\documentclass[journal, twoside]{IEEEtran}
\usepackage{cite}

\ifCLASSINFOpdf
\else
\fi
\usepackage[cmex10]{amsmath}
\usepackage{algorithmic}
  \usepackage[caption=false,font=footnotesize]{subfig}
\usepackage{url}

\usepackage{bm}
\usepackage[pdftex]{graphicx}

\usepackage{amsfonts}
\usepackage[psamsfonts]{amssymb}

\begin{document}
%
\title{Multi-Dimensional Wireless Tomography with Tensor-Based Compressed Sensing}
%
%
%

\author{Kazushi~Takemoto,~\IEEEmembership{Non-Member,~IEEE,}
        Takahiro~Matsuda,~\IEEEmembership{Member,~IEEE,}
	Shinsuke~Hara,~\IEEEmembership{Member,~IEEE,}
	Kenichi~Takizawa,~\IEEEmembership{Member,~IEEE,}
	Fumie~Ono,~\IEEEmembership{Member,~IEEE,}
	and~Ryu~Miura,~\IEEEmembership{Member,~IEEE,}
\thanks{This research was supported in part by 
JSPS Grant-in-Aid for Scientific Research (B)~(Grant No. 25289115) and
Grant-in-Aid for Scientific Research (C)~(Grant No. 25330102). } 
\thanks{K. Takemoto is with Graduate School of Engineering, Osaka
        University, Suita, Osaka, 5650871, Japan~(e-mail:
        k-takemoto@post.comm.eng.osaka-u.ac.jp). }
\thanks{T. Matsuda is with Graduate School of Engineering, Osaka
        University, Suita, Osaka, 5650871, Japan, and  
	National Institute of Information and Communications Technology,
	Yokosuka, Kanagawa, 2390847, Japan~(e-mail:
        matsuda@comm.eng.osaka-u.ac.jp).} 
\thanks{S. Hara is with Graduate School of Engineering, Osaka City
        University, Osaka, 5588585, Japan, and  
	National Institute of Information and Communications Technology,
	Yokosuka, Kanagawa, 2390847, Japan~(e-mail:
        hara@info.eng.osaka-cu.ac.jp).} 
\thanks{K. Takizawa, F. Ono, and R. Miura are with 
	National Institute of Information and Communications Technology,
	Yokosuka, Kanagawa, 2390847, Japan~(e-mail:
        \{takizawa, fumie, ryu\}@nict.go.jp).} 


}

\maketitle

\begin{abstract}
 Wireless tomography is a technique for inferring a physical
 environment
 within a monitored region by analyzing RF signals
 traversed
 across the region.
 In this paper, we consider wireless tomography in a
 two and higher dimensionally structured monitored region, and
 propose a multi-dimensional wireless tomography scheme
based on compressed
 sensing to estimate a spatial distribution of shadowing
 loss in the
 monitored region.
 In order to estimate the spatial
 distribution, we consider two compressed
 sensing frameworks: {\em
 vector-based compressed sensing}
 and {\em tensor-based compressed
 sensing}.
 When the shadowing loss has a high spatial correlation
 in the monitored
 region, the spatial distribution has a sparsity
 in its frequency domain.
 Existing wireless tomography schemes are based on
 the vector-based
 compressed sensing and estimates the distribution
 by utilizing the
 sparsity.
 On the other hand, the proposed scheme is based on
 the tensor-based compressed sensing, which estimates the
 distribution by 
 utilizing its low-rank property.
 We reveal that
 the tensor-based compressed
 sensing has a potential for highly accurate
 estimation as compared with
 the vector-based compressed sensing.

\end{abstract}


%
\IEEEpeerreviewmaketitle

\section{Introduction}

\IEEEPARstart{W}{ireless tomography}
\cite{Kanso2009,Mostofi2011,Patwari2008,Wilson2010}, sometimes called {\em
RF~(Radio Frequency)
tomography} or {\em Radio Tomographic Imaging}, is a 
technique for inferring a physical environment
within a region by
analyzing wireless signals traversed
between wireless
nodes, and it can be used for inferring the locations 
of obstructions from the outside of the region. 
Wireless tomography may lead to developing
building monitoring systems
for security applications and an emergency use
by fire-fighters and
police-officers~\cite{Patwari2008}. 

In general,
wireless signal propagating on a wireless link
loses its power due to
distance, shadowing and multipath fading.
Wireless tomography aims at
estimating a spatial distribution of the shadowing loss
in a monitored
region from a measured power of wireless
signals.
In~\cite{Agrawal2009,Patwari2008}, a spatially correlated
shadowing model is presented, and the power attenuation
of wireless signals is represented as a system of linear
equations of
the spatial distribution.
In~\cite{Patwari2008,Wilson2010},
a regularized weighted
least-squared~(WLS) error estimator is
used to estimate the spatial
distribution. 

{\em Compressed sensing}~\cite{Candes2008,Hayashi2013},
a new paradigm
in signal/image processing, is a promising technique
for wireless
tomography.
By means of compressed sensing, we can solve
underdetermined
linear inverse problems, that is, we can reconstruct
an unknown vector
from fewer measurements than the length of the unknown
vector.
Compressed sensing utilizes the sparsity of the unknown vector,
where most of its elements are exactly or approximately zero.
When the
shadowing loss has a high spatial correlation in the monitored
region, the
spatial distribution has a sparsity in its frequency domain,
which
enables us to naturally apply compressed
sensing.
In~\cite{Kanso2009,Mostofi2011}, compressed sensing-based
wireless tomography schemes are proposed.
In~\cite{Kanso2009}, a
compressed sensing-based wireless tomography is proposed
by extending
the basic idea of wireless tomography
in~\cite{Patwari2008}.
Mostofi~\cite{Mostofi2011} proposes compressive
cooperative mapping in mobile networks,
where mobile nodes collect
measurements and estimate a map of spatial variations
of the parameters
of interest by means of compressed sensing. 

Compressed sensing can be categorized into
{\em vector-based compressed
sensing}~\cite{Beck2009,Candes2008,Hayashi2013,Zibulevsky2010}, {\em 
matrix-based compressed sensing}~\cite{Recht2010,Toh2010}, and {\em
tensor-based compressed sensing}~\cite{Caiafa2013,Gandy2011}. 
Wireless tomography schemes explained in the above
are based on
vector-based compressed sensing and aim at estimating
a spatial
distribution of the shadowing loss in two-dimensional
monitored regions.
In this paper, we try to generalize the wireless tomography
framework in
order to estimate the spatial distribution in higher
dimensional regions and 
propose a {\em multi-dimensional wireless tomography}
scheme using tensor-based compressed sensing. 

Tensors are a higher order generalization of vectors and
matrices~\cite{Gandy2011,Kolda2009,Chen2009,Lathauwer2000},
that is,
vectors and matrices correspond to the 1st-order tensors and the
2nd-order tensors, respectively. 
Tensor-based compressed sensing is a generalization of
matrix-based compressed sensing, and tensor-based and
matrix-based compressed sensing utilize
the low-rank property of unknown
tensors and matrices, respectively,
while vector-based compressed sensing
utilizes the sparsity of unknown vectors.
To the best of our knowledge,
tensor-based compressed sensing has not been applied
to wireless
tomography.
The most important contribution of this paper is to reveal
that the tensor-based compressed sensing has a potential for highly
accurate estimation in multi-dimensional wireless tomography. 

The remainder of this paper is organized as follows. In
section~\ref{sec:cs}, we explain vector, matrix, and tensor-based
compressed sensing.
In section~\ref{sec:problem}, we describe the
problem formulation in multi-dimensional wireless tomography.
In
section~\ref{sec:tomography}, we
propose the multi-dimensional wireless tomography scheme
based on
tensor-based compressed sensing.
In section~\ref{sec:evaluation}, we
evaluate the proposed 
scheme with simulation experiments and discuss
its estimation accuracy 
by comparing it with a wireless tomography scheme based on vector-based
compressed sensing. Finally, we conclude this paper in
section~\ref{sec:conclusion}.

\section{Compressed Sensing}
\label{sec:cs}

As described in the previous section, we consider three frameworks for
compressed sensing:
vector-based compressed sensing, matrix-based compressed
sensing, and
tensor-based compressed sensing,
and hereafter, we refer to them as {\em
vector recovery}, {\em matrix recovery},
and {\em tensor recovery},
respectively. 

\subsection{Vector Recovery} \label{vector}
We first describe the vector recovery,
which is a basic problem in
compressed sensing~\cite{Candes2008,Hayashi2013}. 
In this problem, we consider estimating 
an unknown vector
$\bm{x} = (x_1, \ldots, x_{N})^{\top}\in {\cal R}^{N
\times 1}$ in a linear inverse problem: 
\begin{equation*}
 \bm{y}=\bm{Ax},
\end{equation*}
where $\top$ denotes the transpose operator,
and $\bm{y} \in {\cal R}^{M
\times 1}$ and $\bm{A} \in {\cal R}^{M 
\times N}$ denote a {\em measurement vector}
and a {\em sensing matrix},
respectively, and we assume $M < N$,
that is, an underdetermined linear
system. We also assume that $\bm{A}$ is 
known exactly and $\bm{x}$ is sparse
in some orthonormal basis
$\bm{\Phi}=(\bm{\phi}_1\ \bm{\phi}_2\ \cdots \ \bm{\phi}_N) \in
\mathcal{R}^{N \times N}$ as $\bm{x}
= \bm{\Phi s}$, where $\bm{s} = (s_1~s_2~\cdots~s_N)^\top \in
\mathcal{R}^{N \times 1}$, and $\bm{\phi}_n \in \mathcal{R}^{N \times
1}$~($n = 1, 2, \ldots, N$)\footnote{$s_n$~($n = 1, 2, \ldots, N$) can
be defined over complex number field $\mathcal{C}$
when $\bm{\Phi}$ is
defined by a complex number matrix such as
an inverse Fourier transform
matrix. In this paper, however, we define $s_n$ as a real number by
using an inverse DCT~(Discrete Cosine Transform) matrix in order to simplify the
description. }.
We then have
\begin{equation*}
 \bm{y}=\bm{Ax} =\bm{A \Phi s}.
\end{equation*}

A straightforward approach to the vector recovery is $\ell_0$
optimization: 
\begin{equation*}
 \hat{\bm{s}} = \mathop{\arg\min}_{\bm{s}}\|\bm{s}\|_0 \ \ \ {\rm
  subject\ to}\ \  \bm{y}=\bm{A\Phi s},
\end{equation*}
where $\|\bm{s}\|_0$ is the $\ell_0$ norm
of $\bm{s}$ defined as the
number of nonzero elements in $\bm{s}$.
Finally,
we have an estimate $\hat{\bm{x}}$ by $\hat{\bm{x}}
=\bm{\Phi}\hat{\bm{s}}$.

Because $\ell_0$ norm has the discrete and non-convex natures, 
the above $\ell_0$ optimization problem is difficult to
solve in general. Therefore, in compressed sensing, 
a convex relaxation of $\ell_0$ optimization to $\ell_1$ optimization is
used:
\begin{equation*}
 \hat{\bm{s}} = \mathop{\arg\min}_{\bm{s}}\|\bm{s}\|_1 \ \ \ {\rm
  subject\ to}\ \  \bm{y}=\bm{A\Phi s},
\end{equation*}
where $\|\bm{s}\|_1$ is the $\ell_1$ norm of $\bm{s}$.
Here,
for $\bm{z}  = (z_1~z_2~\cdots~z_L)^\top \in {\cal R}^{L \times 1}$ and $p \geq 1$,
$\ell_p$ norm $\|\bm{z}\|_p$ of $\bm{z}$
is defined as 
\[
 \|\bm{z}\|_p = \left(\sum_{i=1}^{L}|z_i|^p\right)^{1/p}.
\] 

When the measurements are noisy, we can also consider
other optimization 
problems such as $\ell_1$-$\ell_2$ optimization~\cite{Zibulevsky2010}:
\begin{equation}
 \hat{\bm{s}} =
  \mathop{\arg\min}_{\bm{s}}{\left(\frac{1}{2}\|\bm{y}-\bm{A\Phi s}\|_2^2 + \lambda \|\bm{s}\|_1\right)},
\label{FISTA}
\end{equation}
where $\|\bm{y}-\bm{Ax}\|_2$ is the $\ell_2$ norm (Euclidean norm) of
$\bm{y}-\bm{Ax}$ and $\lambda$~($\lambda > 0$) is a regularization
parameter. 
Several algorithms to solve the $\ell_1$-$\ell_2$ optimization have been
proposed, e.g., {\em fast iterative shrinkage-thresholding algorithm}
(FISTA)~\cite{Beck2009,Zibulevsky2010}. 

\subsection{Matrix Recovery} \label{matrix}

In the matrix recovery problem~\cite{Recht2010},
we consider the
following linear inverse problem for an unknown matrix $\bm{X} \in
\mathcal{R}^{N_1\times N_2}$: 
\[
 \bm{y} = \mathcal{A}(\bm{X}),  
\]
where $\mathcal{A}(\cdot)$ represents a linear map $\mathcal{A}:
\mathcal{R}^{N_1 \times N_2} \rightarrow \mathcal{R}^{M \times 1}$ to
obtain the
measurement vector $\bm{y} \in \mathcal{R}^{M \times 1}$. 
We assume that $M < N_1 N_2$, that is,
an underdetermined system.

In the matrix recovery, $\bm{X}$ is estimated by
means of a rank
minimization problem:
\begin{equation}
 \hat{\bm{X}} = \mathop{\arg\min}_{\bm{X}} {\rm rank}(\bm{X})\ \ \ 
\mbox{subject to}\ \ \bm{y}={\cal A}(\bm{X}). \label{rank min}
\end{equation}
Because ${\rm rank}(\bm{X})$ also has the discrete
and non-convex
natures as $\ell_0$ norm, the above rank minimization
problem is
difficult to solve.
In the matrix recovery, therefore, a convex
relaxation of the rank minimization problem 
to the nuclear norm
minimization is used: 
\begin{equation*}
 \hat{\bm{X}} = \mathop{\arg\min}_{\bm{X}}\|\bm{X}\|_\ast \ \ \ 
\mbox{subject to}\ \ \bm{y}={\cal A}(\bm{X}),
\end{equation*}
where $\|\bm{X}\|_*$ is the nuclear norm
of $\bm{X}$, which is defined as 
the sum of its singular values
$\sigma_i$~($i = 1, 2, \ldots, \mathrm{rank}(\bm{X})$):
\begin{equation*}
 \|\bm{X}\|_* = \sum_{i=1}^{{\rm rank}(\bm{X})}\sigma_i.
\end{equation*}
Because all the singular values are nonnegative,
the nuclear norm is
equal to the $\ell_1$ norm of the vector
composed of singular values~\cite{Recht2010}. 

When the measurements are noisy, we can estimate
an unknown matrix $\bm{X}$
as the nuclear norm regularized linear least squares 
problem~\cite{Toh2010}:
\begin{equation}
 \hat{\bm{X}} = \mathop{\arg\min}_{\bm{X}}
{\left(\frac{1}{2}\|\bm{y}-{\cal A}(\bm{X})\|_2^2 + \mu \|\bm{X}\|_*\right)},
\label{APG}
\end{equation}
where $\mu$~($\mu > 0$) is a regularization parameter. 
Several algorithms to solve this problem have been proposed, e.g.,  
the {\em accelerated proximal gradient singular value thresholding algorithm} 
(APG)~\cite{Toh2010}.

\subsection{Tensor Recovery} \label{tensor}

\subsubsection{Tensor Rank}
For an integer $D \geq 3$, the $D$-th order tensor
is referred to as a higher-order tensor. 
The vectorization of the $D$-th order tensor $\bm{X} \in {\cal R}^{N_1
\times N_2 \times \cdots \times N_D}$  
is denoted by ${\rm vec}(\bm{X}) \in {\cal R}^{N_1N_2 \cdots N_D
\times 1}$. 
By the vectorization, the tensor element $(k_1,k_2,\ldots,k_D)$ of
$\bm{X}$ is mapped to 
the $l$-th entry of ${\rm vec}(\bm{X})$, where 
\begin{equation*}
 l= \left\{\sum_{i=1}^{D-1}(k_i-1) \left(
 \prod_{j=i+1}^{D} N_j \right) \right\} + k_D.
\end{equation*}
The mode-$n$ matricization~$(n = 1, 2, \ldots, D)$ of
the $D$-th-order
tensor 
$\bm{X} \in {\cal R}^{N_1 \times \cdots \times N_D}$ is denoted by 
$\bm{X}_{(n)} \in {\cal R}^{N_n \times I_n}$, where 
$I_n = \prod_{i=1 \atop i \neq n}^{D}N_i$.
By the mode-$n$ matricization, the tensor
element $(k_1,k_2 ,\ldots ,
k_D)$ of $\bm{X}$ is mapped  
to the matrix element $(k_n, l_n)$ of $\bm{X}_{(n)}$, where
\begin{equation*}
 l_n= \sum_{i=1 \atop i \neq n}^{D}(k_i-1) \left(
 \prod_{j=i+1 \atop j \neq n}^{D + 1} N_j \right),\quad N_{D + 1} = 1.
\end{equation*}
We refer to $\bm{X}_{(n)}$ as the mode-$n$ unfolding.

There are several notions on
the tensor rank~\cite{Kolda2009}.
In this paper, we consider the n-rank of the $D$-th order
tensor $\bm{X}$, 
which is the tuple of the ranks of the mode-$n$ unfoldings~\cite{Lathauwer2000,Gandy2011}:
\begin{equation*}
 {\rm n}\mbox{-}{\rm rank}(\bm{X}) = \left({\rm rank}(\bm{X}_{(1)}),\ldots , {\rm rank}(\bm{X}_{(D)})\right).
\end{equation*}
\subsubsection{Tensor Recovery}
In the tensor recovery problem, we consider the following linear inverse
problem for 
an unknown $D$-th order tensor $\bm{X} \in \mathcal{R}^{N_1\times
N_2\times \cdots N_D}$:
\[
 \bm{y} = \mathcal{A}(\bm{X}),
\]
where $\mathcal{A}(\cdot)$ represents a linear map
$\mathcal{A}: \mathcal{R}^{N_1 \times N_2 \times \cdots N_D}
 \rightarrow \mathcal{R}^{M \times 1}$ to
 obtain the measurement vector $\bm{y} \in
 \mathcal{R}^{M \times 1}$. 

Now, we define $f({\rm n}\mbox{-}{\rm rank}(\bm{X}))$ as $f({\rm
n}\mbox{-}{\rm rank}(\bm{X})) = \sum_{i=1}^{D}{\rm
rank}(\bm{X}_{(i)})$. 
In the tensor recovery, we consider the minimization of 
function $f({\rm n}\mbox{-}{\rm rank}(\bm{X}))$~\cite{Gandy2011}:
\begin{equation*}
 \hat{\bm{X}} = \mathop{\arg\min}_{\bm{X}} f({\rm
  n}\mbox{-}{\rm rank}(\bm{X})) 
\ \ \ \mbox{subject to}\ \ \bm{y}={\cal A}(\bm{X}).
\end{equation*}
Due to the discrete and non-convex nature of the
tensor rank, the following convex relaxation is considered:
\begin{equation*}
 \hat{\bm{X}} = \mathop{\arg\min}_{\bm{X}} \sum_{i=1}^{D}\|\bm{X}_{(i)}\|_* \ \ \ 
{\rm subject\ to}\ \  \bm{y}={\cal A}(\bm{X}).
\end{equation*}
This problem corresponds to the recovery of
compressed tensor data via
the higher-order singular value decomposition (HOSVD),
which is the most
commonly used generalization of the 
matrix SVD to higher-order tensors~\cite{Lathauwer2000,Chen2009}.
Therefore, the tensor recovery is a generalization of
the matrix recovery.

When the measurements are noisy,
we can estimate an unknown tensor $\bm{X}$
with the following unconstrained optimization:
\begin{equation}
 \hat{\bm{X}} = \mathop{\arg\min}_{\bm{X}}
{\left(\frac{\mu}{2}\|\bm{y}-{\cal A}(\bm{X})\|_2^2 + 
 \sum_{i=1}^{D}\|\bm{X}_{(i)}\|_*\right)}, \label{DR-TR}
\end{equation}
where $\mu$~$(\mu > 0)$ is a regularization parameter.
Several algorithms to solve this problem have also been proposed, e.g., 
{\em Douglas-Rachford splitting for tensor recovery} 
(DR-TR)~\cite{Gandy2011}.

\section{Problem Formulation}
\label{sec:problem}

In wireless tomography, nodes inject wireless
signals into a
{\em monitored region}, and characteristics
such as power attenuation
due to obstructions are inferred from the received
wireless signals. 
In this paper,
we consider the $D$-dimensional wireless
tomography~($2 \leq D \leq 4$),
where the monitored region is represented by a $D$-dimensional
structure on Cartesian coordinates.
For the case of $D = 4$,
wireless tomography is described with 3 spatial
axes and the time
axis.
Here, 
we discretize the time axis into intervals with the same time
unit and let $t$~($t = 1, 2, \ldots$) denote the $t$-th time interval,
which is also referred to as time $t$ hereafter.

\begin{figure}[!t]
\centering
\subfloat[]{\includegraphics[scale=0.4,clip]{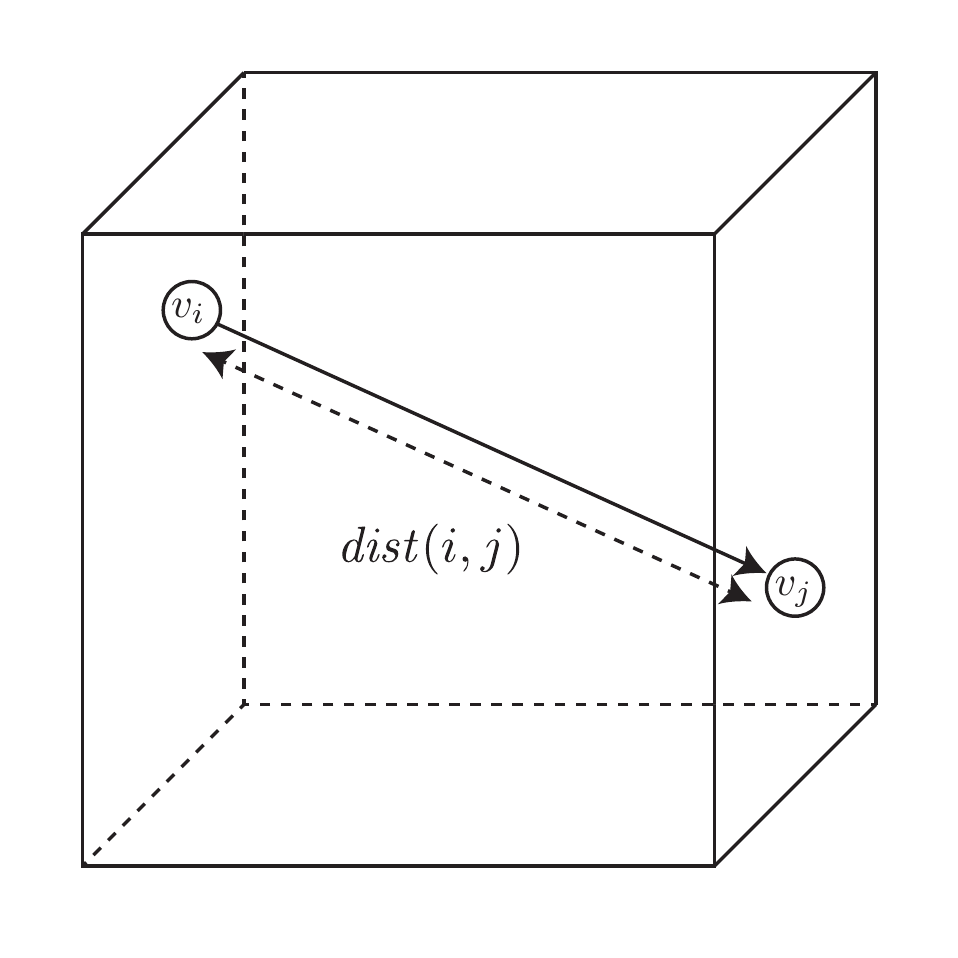}\label{subfig:pathloss}} \\
\subfloat[]{\includegraphics[scale=0.4,clip]{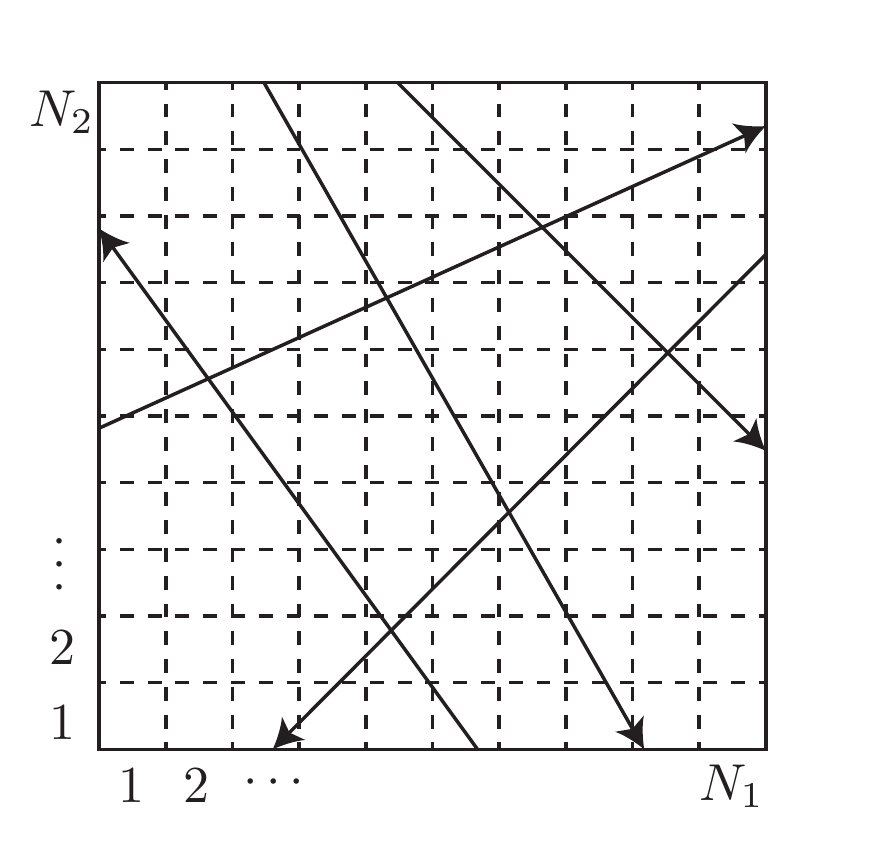}\label{subfig:wirelesstomography}}\hspace{20pt}
\subfloat[]{\includegraphics[scale=0.4,clip]{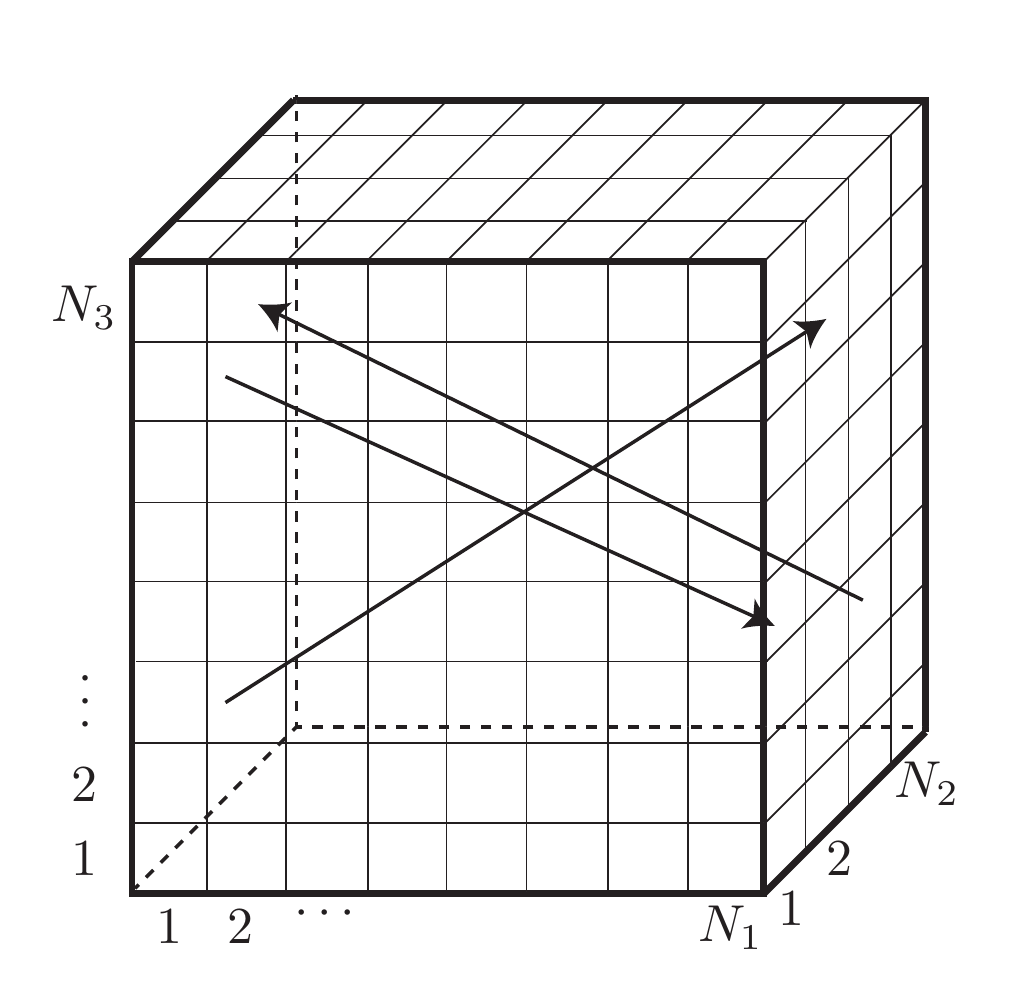}\label{subfig:wirelesstomography3D}}
\caption{Multi-dimensional wireless tomography. (a) Wireless nodes are
 deployed on the border of the monitored region. (b) The monitored
 region for 2-dimensional wireless tomography. (c)
 The monitored region for 3-dimensional wireless tomography.}
\label{fig:wirelesstomography}
\end{figure}

The path loss of signal propagated on a wireless link 
consists of the large-scale path loss due to distance,
shadowing
loss due to obstructions, and non-shadowing loss due to multipath
fading~\cite{Hashemi1993, Agrawal2009, Patwari2008}. 
Let ${\cal V}$ denote a set of wireless nodes,
which are deployed on
the border of the monitored region as shown in
Fig.~\ref{subfig:pathloss}. 
Suppose that a wireless signal is transmitted from a transmitter $v_i
\in \cal{V}$ to a receiver $v_j \in \cal{V}$~($i, j=1, 2,\ldots,|{\cal
V}|,\ j \neq i$). 
We define ${\it dist}(i,j)$ and $\bar{P}({\it dist}(i,j))$ as the
distance and the large-scale path loss in dB between $v_i$ and $v_j$,
respectively.
In this case, we
can model the received signal power $P_{i,j, t}$~[dBm]
at $v_j$ 
observed at time $t$ as  
\begin{eqnarray*}
 P_{i,j, t} &=& P_{\rm TX} - \bar{P}({\it dist}(i,j)) - Z_{i,j, t},\\
 Z_{i,j, t} &=& Z^{(1)}_{i,j, t} + Z^{(2)}_{i,j, t},
\end{eqnarray*}
where $P_{\rm TX}$, $Z^{(1)}_{i,j, t}$ and
$Z^{(2)}_{i,j, t}$ denote the
transmitted power at $v_i$  
in dBm, the shadowing loss in dB, and 
the non-shadow fading loss in dB, respectively.
Furthermore,
$\bar{P}({\it dist}(i,j))$ is given by 
\begin{equation*}
 \bar{P}({\it dist}(i,j)) = 10\alpha \log({\it dist}(i,j)) +\beta,
\end{equation*}
where $\alpha$ and $\beta$ are constants and $\alpha \geq
2$~\cite{Hashemi1993}.
Using the line integral over the wireless link ${\it path}(i, j)$
between
$v_i$ and $v_j$, we have 
\begin{equation*}
 Z^{(1)}_{i, j, t}= \int_{{\it path}(i,j)} g(\bm{r}, t) d\bm{r},
\end{equation*}
where $\bm{r} \in \mathcal{R}^{3}$ denotes a coordinate in
the monitored region and 
$g(\bm{r}, t)$~[dB/m] denotes the power attenuation due to
the shadowing loss on location $\bm{r}$ at time $t$~\cite{Agrawal2009,
Mostofi2011,Patwari2008}. 
Note that $g(\bm{r},t)=0$ if there is no obstruction on $\bm{r}$.
For $Z^{(2)}_{i,j, t}$,
we assume a wide-sense
stationary Gaussian
process with zero mean and variance $\eta^2$.

\begin{figure}[!t]
 \centering
 \includegraphics[scale=0.6]{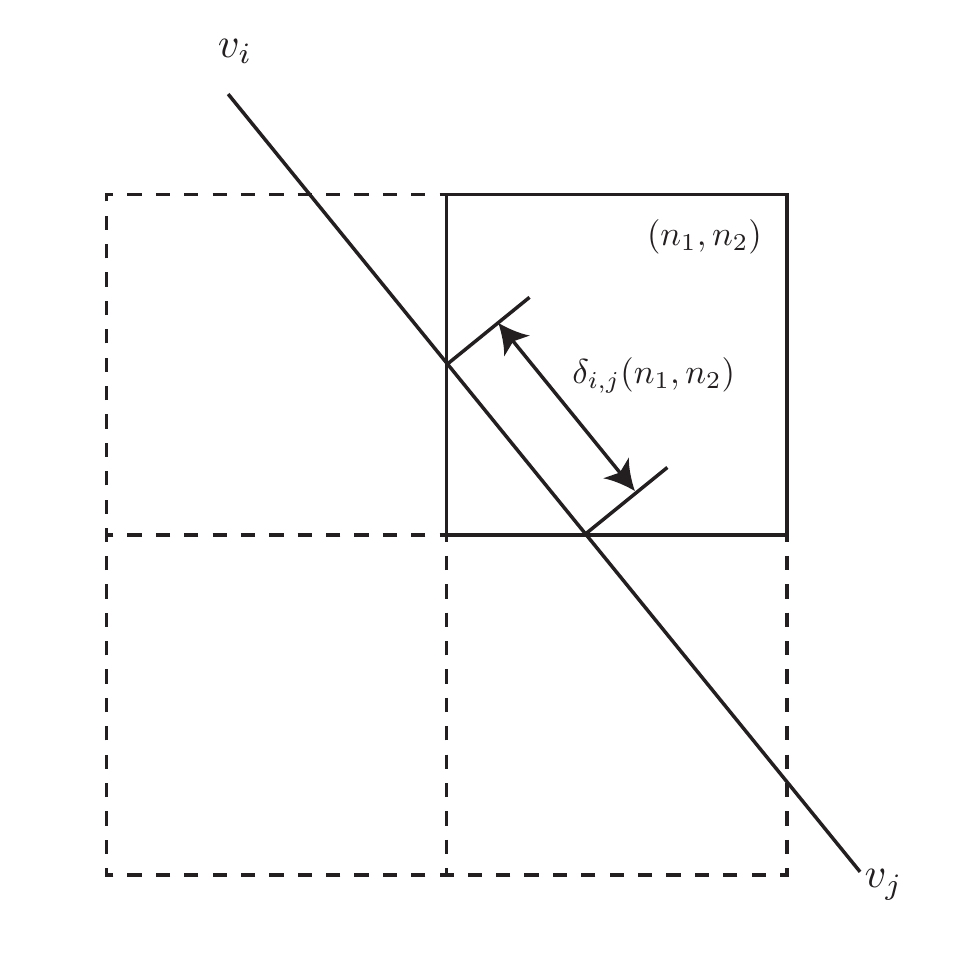}
 \caption{An example of overlap distance $\delta_{i, j}(n_1, n_2,
 \ldots, n_D)$ on wireless link $\mathit{path}(i, j)$ between wireless
 nodes $v_i \in \cal{V}$ and $v_j \in \cal{V}$ for $D = 2$. }
 \label{fig:intersect}
\end{figure}

Let us
divide the monitored region into $3$-dimensional voxels 
$(n_1, n_2, n_3)\ (n_i = 1,2, \ldots, N_i,\ i=1,2,3)$, and represent
$\Delta(n_1, n_2, n_3) \subset \mathcal{R}^3$ as a subset of the
monitored region within voxel $(n_1, n_2, n_3)$. 
Here, 
we assume that $g(\bm{r} \in \Delta(n_1, n_2, n_3), t = n_4)$~($n_i = 1,
2, \ldots, N_i$, $i = 1, 2, 3, 4$) has a constant value $X_{n_1, n_2, n_3, n_4}$  
within voxel $(n_1, n_2, n_3)$.
Figs.~\ref{subfig:wirelesstomography}
and~\ref{subfig:wirelesstomography3D} show examples of a monitored 
region divided into voxels for $2$-dimensional wireless tomography and
$3$-dimensional wireless tomography, respectively. 
We then have
\begin{equation*}
 Z^{(1)}_{i,j, n_4}=\sum_{n_1, n_2, n_3} 
 \delta_{i,j}(n_1, n_2, n_3) X_{n_1, n_2 ,n_3, n_4},
\end{equation*}
where $\delta_{i,j}(n_1, n_2, n_3)$ is the overlapped
distance between
wireless link ${\it path}(i, j)$ and voxel $(n_1, n_2, n_3)$~(See
Fig.~\ref{fig:intersect}). 
Note that $\delta_{i,j}(n_1, n_2, n_3) = 0$ if ${\it path}(i, j)$ 
does not traverse voxel~$(n_1, n_2, n_3)$.

Now, let $\mathcal{Q}_{n_4} = \{(v_i, v_j) \mid v_i, v_j \in
\mathcal{V}, \ i, j \in \{1, 2, \ldots, |\mathcal{V}|\}\}$~($n_4 =
1, 2, \ldots, N_4$) denote a set of pairs of nodes used for
measurements at time $n_4$. 
In addition,
let $(v_{i_m}^{(n_4)}, v_{j_m}^{(n_4)}) \in \mathcal{Q}_{n_4}$~($m = 1,
2, \ldots, M_{n_4}$, $i_m, j_m \in \{1, 2, \ldots, |\mathcal{V}|\}$, $i_m
\neq j_m$) denote $M_{n_4}$ pairs of wireless nodes, where $M_{n_4} = |\mathcal{Q}_{n_4}|$. 
Given $P_{i_m,j_m, n_4}$, $P_{\rm TX}$ and $\bar{P}({\it
dist}({i_m,j_m}))$,  
we can obtain the following linear equation:
\begin{eqnarray}
 y_m^{(n_4)} &\triangleq& 
 P_{\rm TX} - P_{i_m,j_m, n_t} - \bar{P}({\it dist}({i_m,j_m})) \nonumber \\
\nonumber &=& \sum_{n_1, n_2, n_3}\delta_{i_m,j_m}(n_1, n_2, n_3)X_{n_1, n_2,
  n_3, n_4} \\
&& + Z^{(2)}_{i_m,j_m, n_4}, 
    \label{eqn:linear}
\end{eqnarray}
where $y_m^{(n_4)}$ is referred to as the $m$-th {\em measured
shadowing loss} at time $n_4$. 
Furthermore,
let $\bm{y}^{(n_4)} =
(y_{1}^{(n_4)}~y_{2}^{(n_4)}~\cdots~y_{M_{n_4}}^{(n_4)})^\top \in
\mathcal{R}^{N_1N_2N_3 \times 1}$ denote a
measurement vector at time $n_4$ and $\bm{X}^{(n_4)} = \{X_{n_1, n_2,
n_3, n_4} \mid n_i = 1, 2, \ldots, N_i, i = 1, 2, 3\}$ denote a {\em loss
field tensor} at time $n_4$.  With a linear map
$\mathcal{A}^{(n_4)}: \mathcal{R}^{N_1\times N_2 \times N_3} \rightarrow
\mathcal{R}^{M_{n_4} \times 1}$,
(\ref{eqn:linear}) is  rewritten as follows:
\[
 \bm{y}^{(n_4)} = \mathcal{A}^{(n_4)}(\bm{X}^{(n_4)}) + \bm{w}^{(n_4)},
\]
where $\bm{w}^{(n_4)} = (Z_{i_1, j_1, n_4}^{(2)}~Z_{i_2,
j_2, n_4}^{(2)}~\cdots~Z_{i_{M_{n_4}}, 
j_{M_{n_4}}, n_4}^{(2)})$.

Defining measurement vector $\bm{y}$ as
\begin{eqnarray}
\nonumber
 \bm{y} &\triangleq& \left(
 \begin{array}{c}
  \bm{y}^{(1)}\\
  \bm{y}^{(2)}\\
  \vdots \\
  \bm{y}^{(N_4)}\\
 \end{array}
 \right) \\
\label{eqn:linear2}
&=& 
 \left(
 \begin{array}{c}
  \mathcal{A}^{(1)}(\bm{X}^{(1)})\\
  \mathcal{A}^{(2)}(\bm{X}^{(2)})\\
  \vdots \\
  \mathcal{A}^{(N_4)}(\bm{X}^{(N_4)})\\
 \end{array}
 \right) + 
 \left(
 \begin{array}{c}
  \bm{w}^{(1)}\\
  \bm{w}^{(2)}\\
  \vdots \\
  \bm{w}^{(N_4)}\\
 \end{array}
 \right),
\end{eqnarray}
we finally
reformulate (\ref{eqn:linear2}) with a linear map ${\cal A} : {\cal
R}^{N_1 \times N_2 \times N_3 \times N_4} \rightarrow {\cal R}^{M \times
1}$:
\begin{equation}
 \bm{y} = {\cal A}(\bm{X}) + \bm{w} \label{tensor equation}, 
\end{equation}
where $\bm{X} = \{\bm{X}^{(n_4)} \mid n_4 = 1, 2, \ldots, N_4\} \in
\mathcal{R}^{N_1\times N_2\times N_3\times N_4}$ denote
a loss field tensor and $\bm{w} =
((\bm{w}^{(1)})^\top~(\bm{w}^{(2)})^\top~\cdots~(\bm{w}^{(N_4)})^\top)^\top
\in \mathcal{R}^{N_1N_2N_3N_4 \times 1}$
denote a noise vector.

Wireless tomography is a linear inverse problem
to estimate $\bm{{\rm
X}}$ from the measurement vector $\bm{y}$. 
Note that $D$-dimensional wireless tomography
for $D = 2, 3$ can be
formulated as special cases of the 4-dimensional
wireless tomography.
Namely,
the 2-dimensional wireless tomography
corresponds to the 4-dimensional
wireless tomography with $N_1 > 1$, $N_2> 1$,
and $N_3 = N_4 
= 1$, where the loss field tensor $\bm{X}$
has two spatial axes~(i.e.,
$x$-axis, $y$-axis). For the case of $D = 3$,
we can consider two cases: three spatial
axes~(i.e., $x$-axis, $y$-axis, and $z$-axis),
and two spatial axes and
time axis.
The former corresponds to the 4-dimensional wireless
tomography with $N_1 > 1, N_2 > 1$, $N_3 > 1$,
and $N_4 = 1$, while
the latter corresponds to the 4-dimensional wireless tomography
with  $N_1 > 1,
N_2 > 1, N_3 = 1$, and $N_4 > 1$.

\section{Wireless Tomography with Compressed Sensing}
\label{sec:tomography}

We assume that
$X_{n_1,\ldots,n_D}$ has a high spatial correlation,
which enables us to
estimate $X_{n_1,\ldots,n_D}$ by means of compressed sensing. 
In the following subsections, we first describe
the wireless tomography
scheme based on the vector recovery, which has been studied 
in~\cite{Kanso2009,Mostofi2011}, and then, we explain the
wireless tomography scheme based on the tensor recovery.

\subsection{Vector Recovery-Based Wireless Tomography} \label{estimation_vector} 

We define a {\em loss field vector} $\bm{x}$ as $\bm{x}={\rm 
vec}(\bm{X}) \in {\cal R}^{N_1 N_2 N_3 N_4 \times 1}$, and
reformulate 
(\ref{tensor equation}) as $\bm{y}=\bm{Ax}+\bm{w}$,
where $\bm{A} \in
{\cal R}^{M \times N_1 N_2 N_3 N_4}$ denotes
a sensing matrix.
Let ${\cal F}$ denote a linear map that transforms
$\bm{X}$ to
its frequency domain representation such as
the $D$-dimensional {\em
discrete Fourier transform} (DFT) and
the $D$-dimensional {\em discrete
cosine transform} (DCT).  
In addition,
let $\bm{S}= {\cal F}(\bm{X}) \in {\cal C}^{N_1
\times N_2 \times N_3 \times N_4}$ and $\bm{s} = \mathrm{vec}(\bm{S})$ denote
the frequency domain representation of $\bm{X}$
and its vectorization,
respectively. 
The matrix that transforms $\bm{s}$ to $\bm{x}$ is
denoted by $\bm{\Phi}
\in {\cal C}^{N_1 \cdots N_D \times N_1 \cdots N_D}$, that is,
$\bm{x}=\bm{\Phi s}$, so we have
\begin{equation*}
 \bm{y}=\bm{A \Phi s}+\bm{w}.
\end{equation*}
Then,
let  $\bm{F}_{N} = \{F_{k, n}\mid 1 \leq k, n \leq N\}$~($N \geq 1$) denote an $N \times N$ unitary matrix
such as the one-dimensional DFT matrix and the one-dimensional DCT
matrix. In the case of DCT, $F_{k, n}$ is given by
\[
 F_{n, k} = 
\left\{
\begin{array}{cl}
\displaystyle \frac{ 1}{\sqrt{N}}, & k = 1, 1 \leq n \leq N,\\
 \displaystyle \sqrt{\frac{2}{N}}\cos\frac{\pi (2n + 1)k}{2N}, &  2 \leq k
  \leq N, 1 \leq n \leq N.
\end{array}
\right.
\]
By using $\bm{F}_N$, $\bm{\Phi}$ can be written as
\[
 \bm{\Phi} = \bm{F}_{N_1}^\top \otimes \bm{F}_{N_2}^\top \otimes
 \bm{F}_{N_3}^\top \otimes \bm{F}_{N_4}^\top, 
\]
where $\otimes$ represent the Kronecker product, and  
if $\bm{B} = \{b_{i, j}\}$ and $\bm{C} = \{c_{k, l}\}$ are $L_1 \times
L_2$ and $L_3 \times L_4$ matrices, respectively, $\bm{B} \otimes
\bm{C}$ is defined as 
\[
 \bm{B} \otimes \bm{C} = 
\left(
\begin{array}{cccc}
b_{1,1}\bm{C} &  b_{1,2}\bm{C} & \cdots  & b_{1,L_2}\bm{C} \\
b_{2,1}\bm{C} &  b_{2,2}\bm{C} & \cdots  & b_{2,L_2}\bm{C} \\
\vdots & \vdots & \ddots & \vdots \\
b_{L_1,1}\bm{C} &  b_{L_1,2}\bm{C} & \cdots  & b_{L_1,L_2}\bm{C} \\
\end{array}
\right).
\]
Consequently, we can obtain an estimate $\hat{\bm{s}}$ of the frequency
domain representation 
$\bm{s}$ by the sparse vector recovery, as explained in
section~\ref{vector}, and then have the estimate
$\hat{\bm{x}}$ of loss field vector $\bm{x}$ by
$\hat{\bm{x}}=\bm{\Phi}\hat{\bm{s}}$.  

\subsection{Tensor Recovery-Based Wireless Tomography}
\label{estimation_tensor}

The wireless tomography scheme based on
the tensor recovery estimates the loss field
tensor $\bm{X}$ from the measurement vector $\bm{y}$ by
means of the
tensor recovery explained in section~\ref{tensor}. 
Because
each element of measurement vector $\bm{y}$
includes a noise in a practical situation,
we estimate $\bm{X}$ by means
of (\ref{DR-TR}).
Namely,
from (\ref{eqn:linear2}) and (\ref{tensor equation}),
we can rewrite
(\ref{DR-TR}) as 
\begin{eqnarray*}
  \hat{\bm{X}} &=& \mathop{\arg\min}_{\bm{X}}
{\left(\frac{\mu}{2}\|\bm{y}-{\cal A}(\bm{X})\|_2^2 + 
 \sum_{i=1}^{D}\|\bm{X}_{(i)}\|_*\right)} \\
&=& \mathop{\arg\min}_{\bm{X}}
\left(\frac{\mu}{2}\sum_{n_4 = 1}^{N_4}\|\bm{y}^{(n_4)}-{\cal
 A}^{(n_4)}(\bm{X}^{(n_4)})\|_2^2 \right. \\
&& \left. +  \sum_{i=1}^{D}\|\bm{X}_{(i)}\|_\ast \right).
\end{eqnarray*}

It is worth mentioning that the
matrix recovery-based wireless tomography
corresponds to the
tensor recovery-based wireless tomography for $D =
2$.
Therefore, in the following,
we use the tensor recovery-based wireless
tomography for $D = 2$ and
the matrix recovery-based wireless tomography
exchangeably.

\section{Performance Evaluation} 
\label{sec:evaluation}
\begin{figure}[!t]
 \centering
 \subfloat[$D = 2$.]{\includegraphics[scale=0.35]{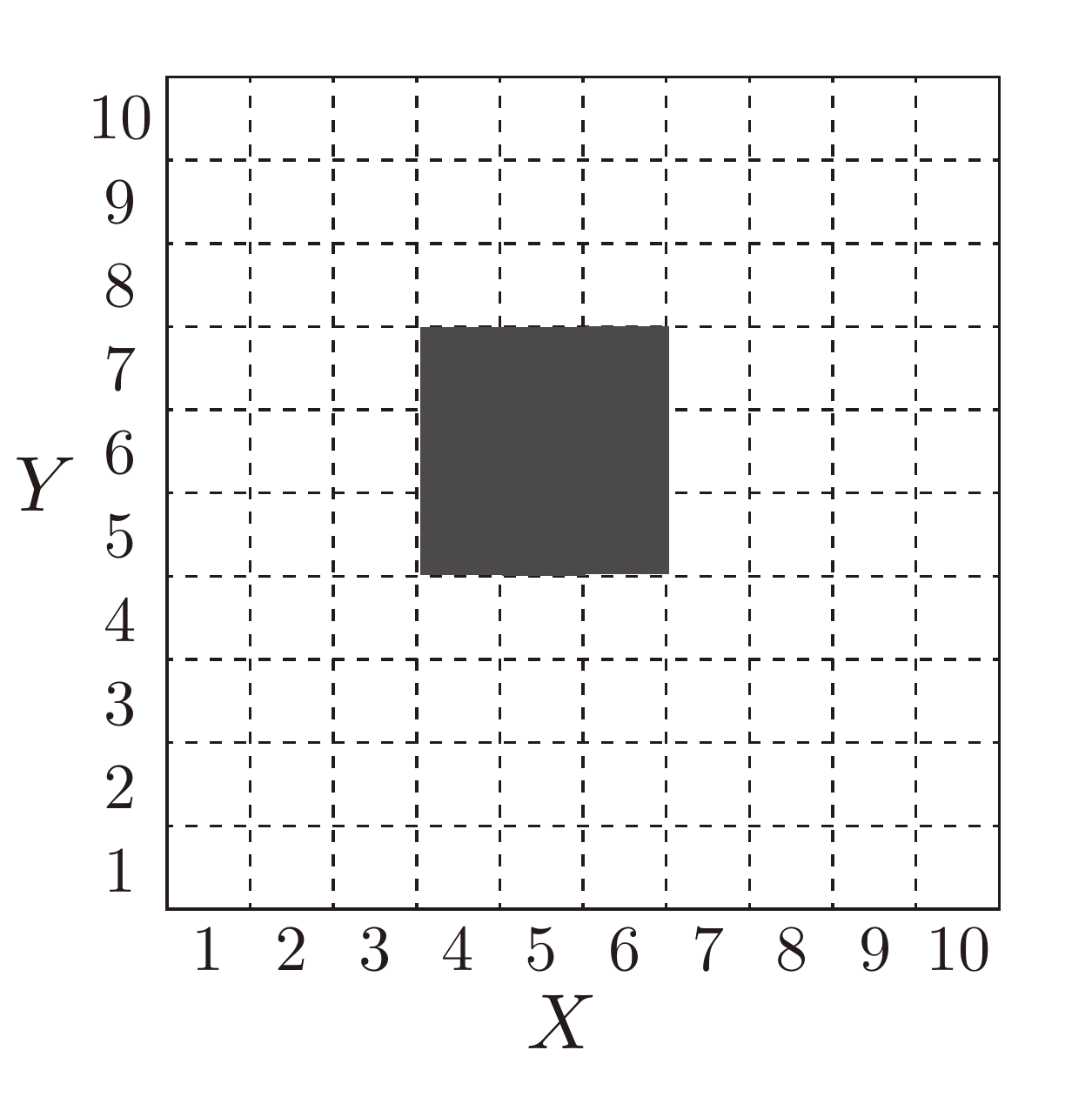}  
 \label{fig:channel2D}} \\
\subfloat[$D = 3$. ]{\includegraphics[scale=0.5]{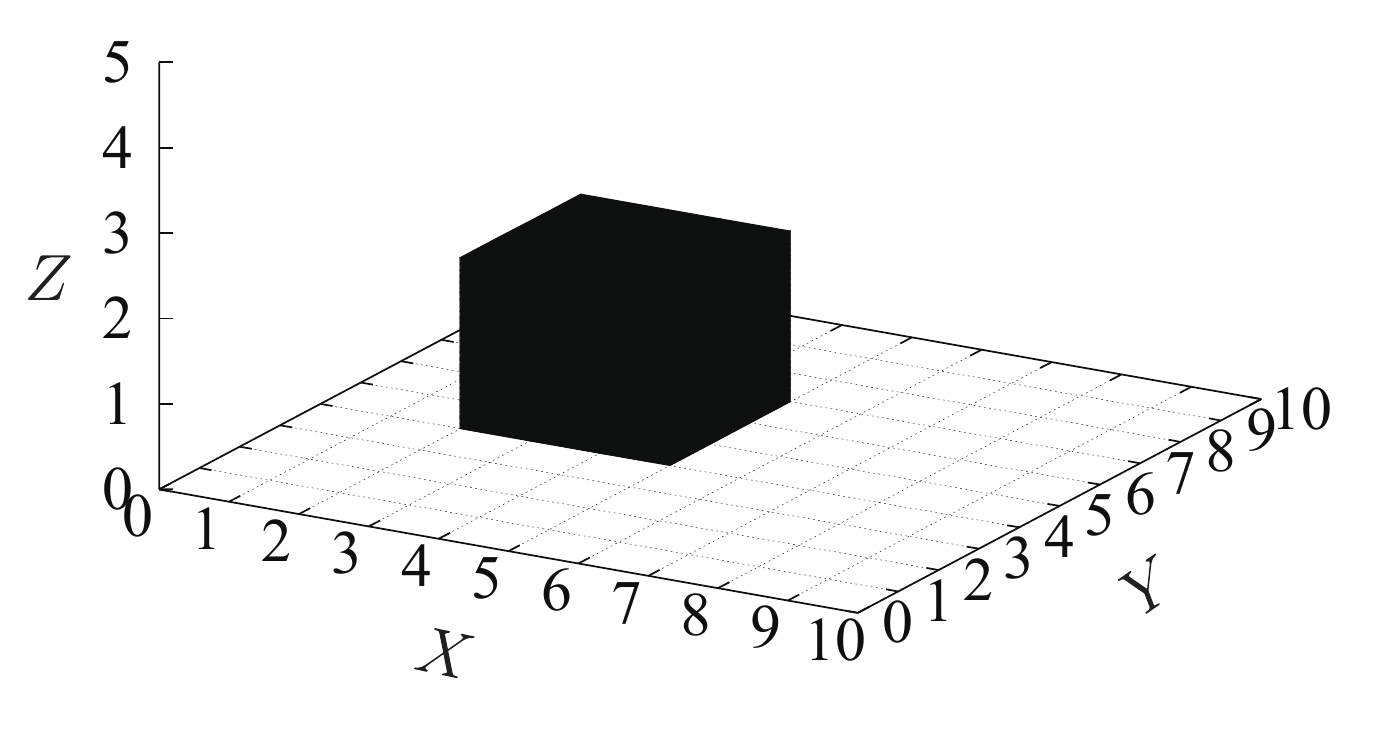}  
 \label{fig:channel3D}} \\
\subfloat[$D = 4$. ]{\includegraphics[scale=0.4]{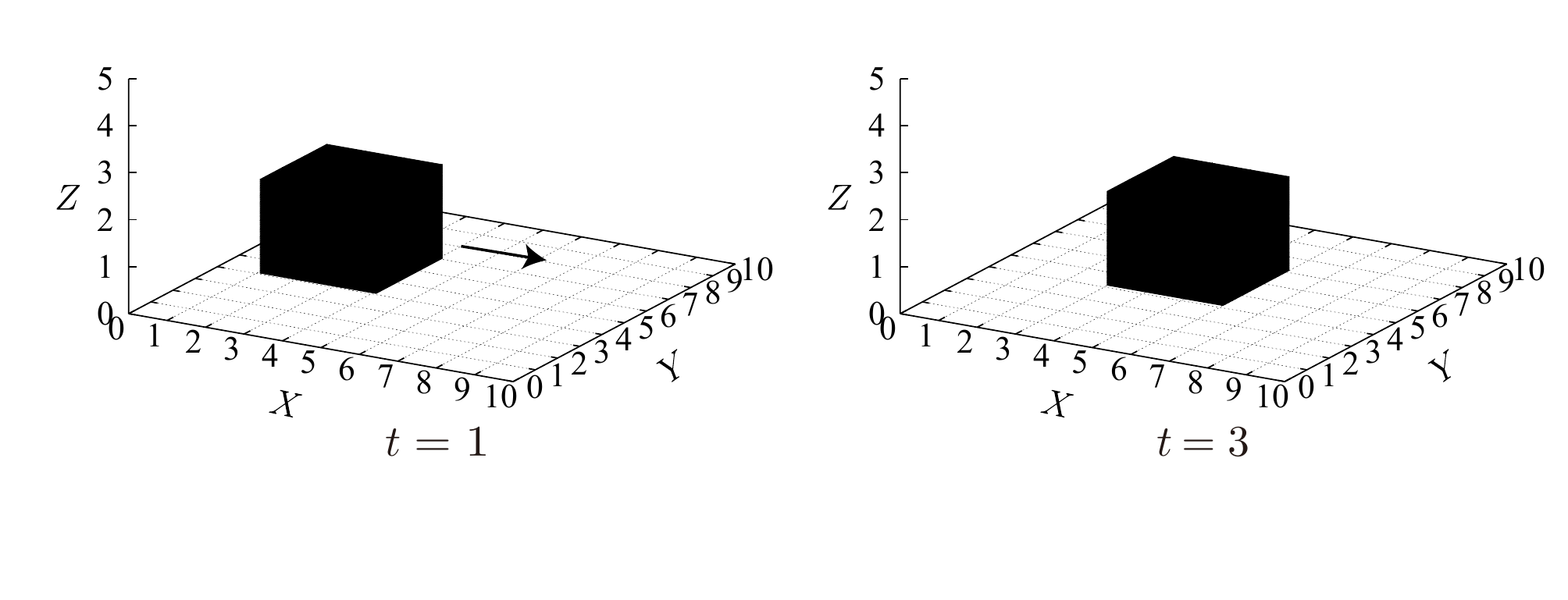}  
 \label{fig:channel4D}}
 \caption{Monitored regions for simulation experiments.}
\end{figure}

\subsection{Simulation Setup}
\begin{figure}[!t]
 \centering
\subfloat[$D =
 2$.]{\includegraphics[scale=0.5]{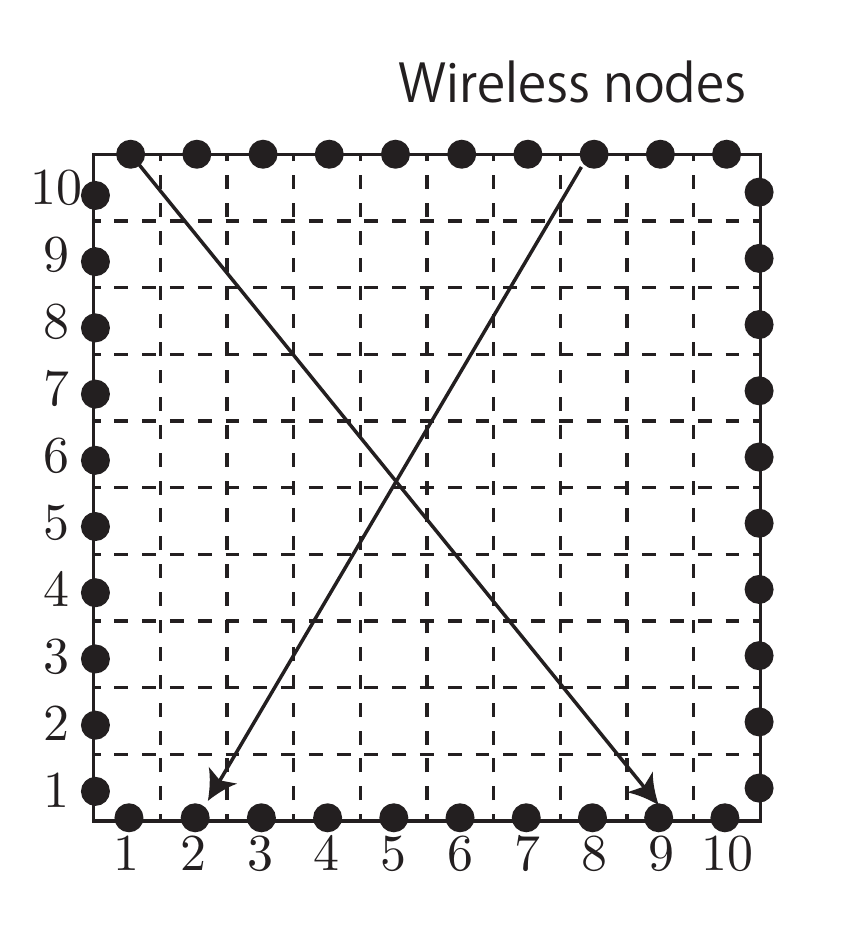}
 \label{fig:2D_nodeplace}} \\ 
\subfloat[$D =
 3, 4$.]{\includegraphics[scale=0.5]{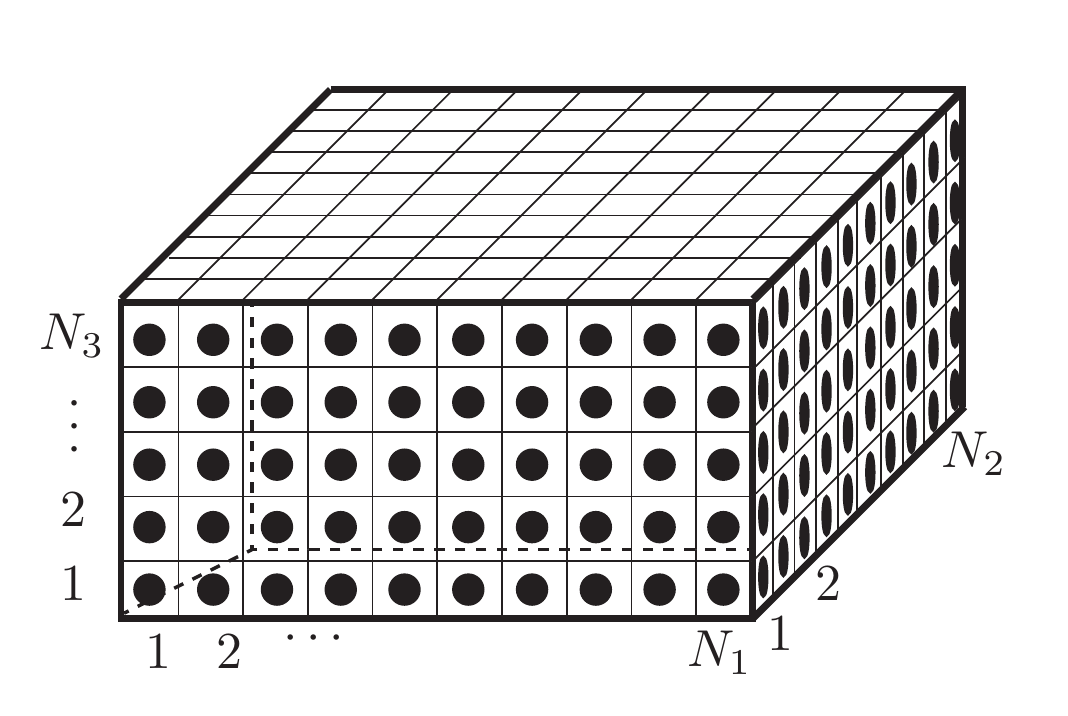}
 \label{fig:3D_nodeplace}} 
 \caption{Node placement in simulation experiments.}
 \label{fig:nodeplace}
\end{figure}

In this section, we demonstrate the performance
of the tensor recovery scheme by
comparing it with that of the vector
recovery scheme for $D = 2, 3, 4$. 
Figs.~\ref{fig:channel2D}, \ref{fig:channel3D},
and \ref{fig:channel4D}
show the monitored regions for $D = 2$, $D = 3$, and $D = 4$, respectively. 
In the case of $D = 2$, we set $N_1 = N_2 = 10$ and
place a $3\times
3$ square-shaped obstruction in the monitored region, which is
represented by dark pixels in Fig.~\ref{fig:channel2D}. 
The elements of $\bm{X} \in \mathcal{R}^{N_1 \times N_2}$
are set to $10$  
within the dark pixels and $0$ within the other pixels. 
40 wireless nodes are placed on the border
of the monitored region as shown
in Fig.~\ref{fig:2D_nodeplace}.
Next,
in the case of $D = 3$, we set $N_1 = N_2 = 10,\ N_3 = 5$
and place a $3 
\times 3 \times 2$ obstruction, which is represented by dark
voxels in Fig.~\ref{fig:channel3D}.
The elements of $\bm{X} \in
\mathcal{R}^{N_1 \times N_2 \times N_3}$
are set to $10$ within the dark
voxels and $0$ within the other pixels.
200 wireless nodes are placed on
the sides of the monitored region as shown 
in Fig.~\ref{fig:3D_nodeplace}.
Finally,
in the case of $D = 4$, we set $N_1 =
N_2 = 10$, $N_3 = 5$, and $N_4 = 3$,
where the monitored region is the
same environment as the case of $D = 3$
and the obstruction is moving as 
shown in Fig.~\ref{fig:channel4D}.

In each simulation experiment,
$M$ pairs of wireless nodes are randomly
chosen to establish $M$ wireless links,
and in each pair, a randomly
chosen node is set to a transmitter and the other is set to a
receiver.
We assume that each measurement is contaminated with
a Gaussian
noise with zero mean and variance $\eta^2$. 

In the vector recovery scheme, we use the DCT
to transform the loss field tensor $\bm{X}$
to its frequency representation,
and estimate the loss field tensor by
solving the optimization problem~(\ref{FISTA}) with
FISTA~\cite{Beck2009,Zibulevsky2010}.
The regularization parameter
$\lambda$ in~(\ref{FISTA}) is set to
$1.0$.
On the other
hand, in the tensor recovery scheme, we estimate the loss field tensor
$\bm{X}$ by
solving the optimization problem~(\ref{APG})~\cite{Toh2010} for $D = 2$
and the optimization problem~(\ref{DR-TR})~\cite{Gandy2011} for $D = 3,
4$.
The regularization parameter $\mu$ is set to $1.0$. 
Note that in this paper, we do not consider the
optimization of the
regularization parameters $\lambda$ and $\mu$,
which is beyond the scope
of the paper. 

We evaluate the performance
of the vector and tensor recovery schemes
in terms of the normalized reconstruction error $\epsilon$
between the true loss field tensors
and the corresponding estimated loss field
tensors. 
In more detail,
for given $M$ and $\eta$,
we conduct $N_{\rm run}$ independent
simulation experiments to calculate $\epsilon$,
which is defined as 
\begin{equation}
  \epsilon = \frac{1}{N_{\rm run}}\sum_{i = 1}^{N_{\rm
  run}}\frac{\|\hat{\bm{X}_{[i]}} 
  -\bm{X} \|_{\rm F}}{\|\bm{X}\|_{\rm F}}.
  \label{eq:hara1}
\end{equation}
In (\ref{eq:hara1}),
$\hat{\bm{X}}_{[i]}$~($i = 1, 2, \ldots, N_{\rm run} $) 
denote
the estimated loss field tensor obtained by
the $i$-th simulation experiment
and
$\|\cdot \|_{\rm F}$ represents the Frobenius norm.
Here, for
$\bm{{\rm Z}} 
= \{Z_{n_1, n_2, \ldots, n_D} \mid n_i = 1, 2, \ldots, N_i, i = 1, 2,
\ldots, D\} \in \mathcal{R}^{N_1 \times N_2 \times \cdots N_D}$, 
its Frobenius norm is defined as
\begin{equation*}
 \|\bm{{\rm Z}}\|_{\rm F}= \left(\sum_{n_1=1}^{N_1}\sum_{n_2=1}^{N_2} \cdots 
 \sum_{n_D=1}^{N_D} Z_{n_1,n_2, \ldots,n_D}^2 \right)^{\frac{1}{2}}.
\end{equation*}

\subsection{Simulation Results} 
\begin{figure}[tt]
 \centering
\subfloat[Vector recovery.]{
 \includegraphics[scale=0.3,clip]{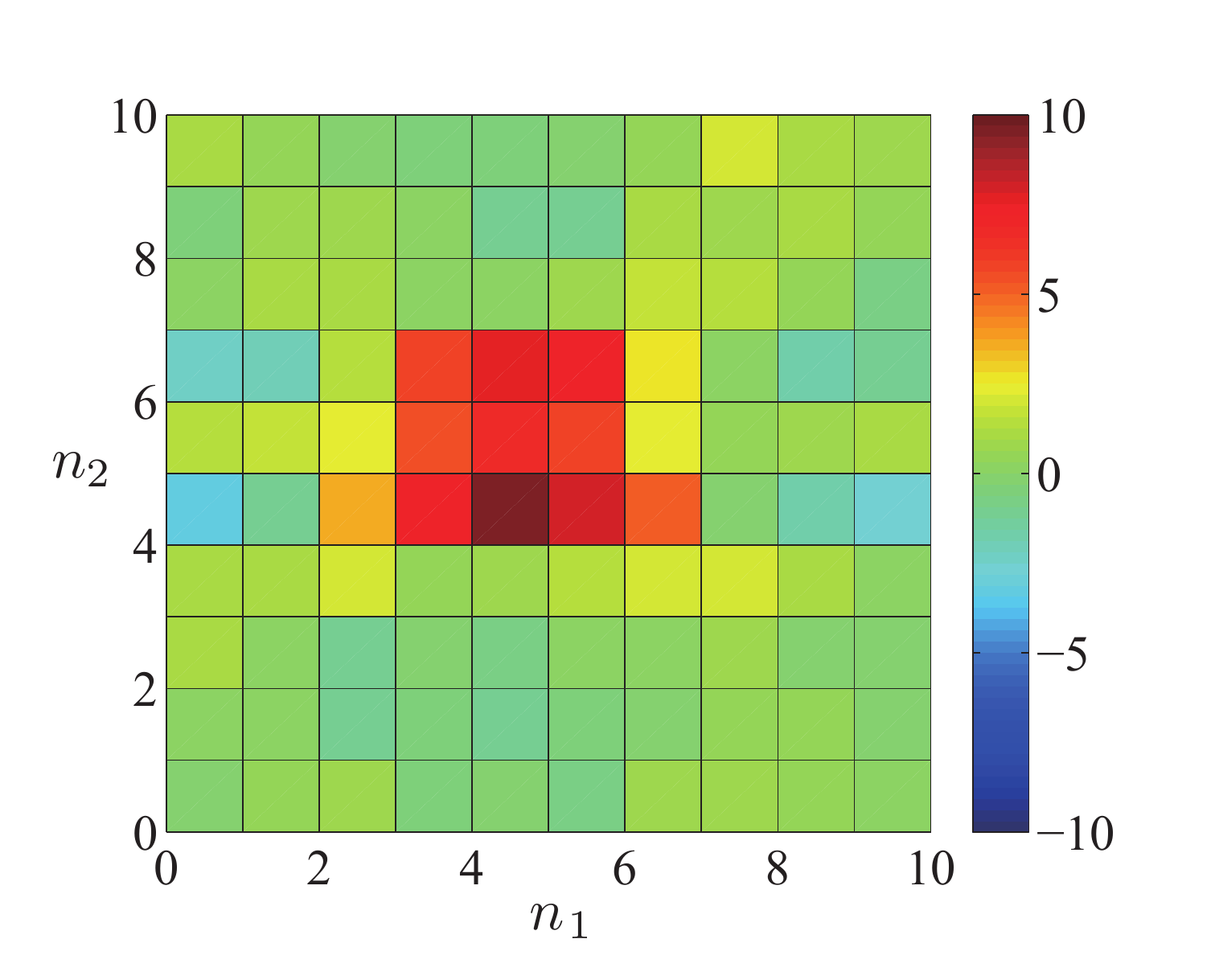} 
 \label{fig:2DDCT}
 }
\subfloat[Matrix recovery.]{
 \includegraphics[scale=0.3,clip]{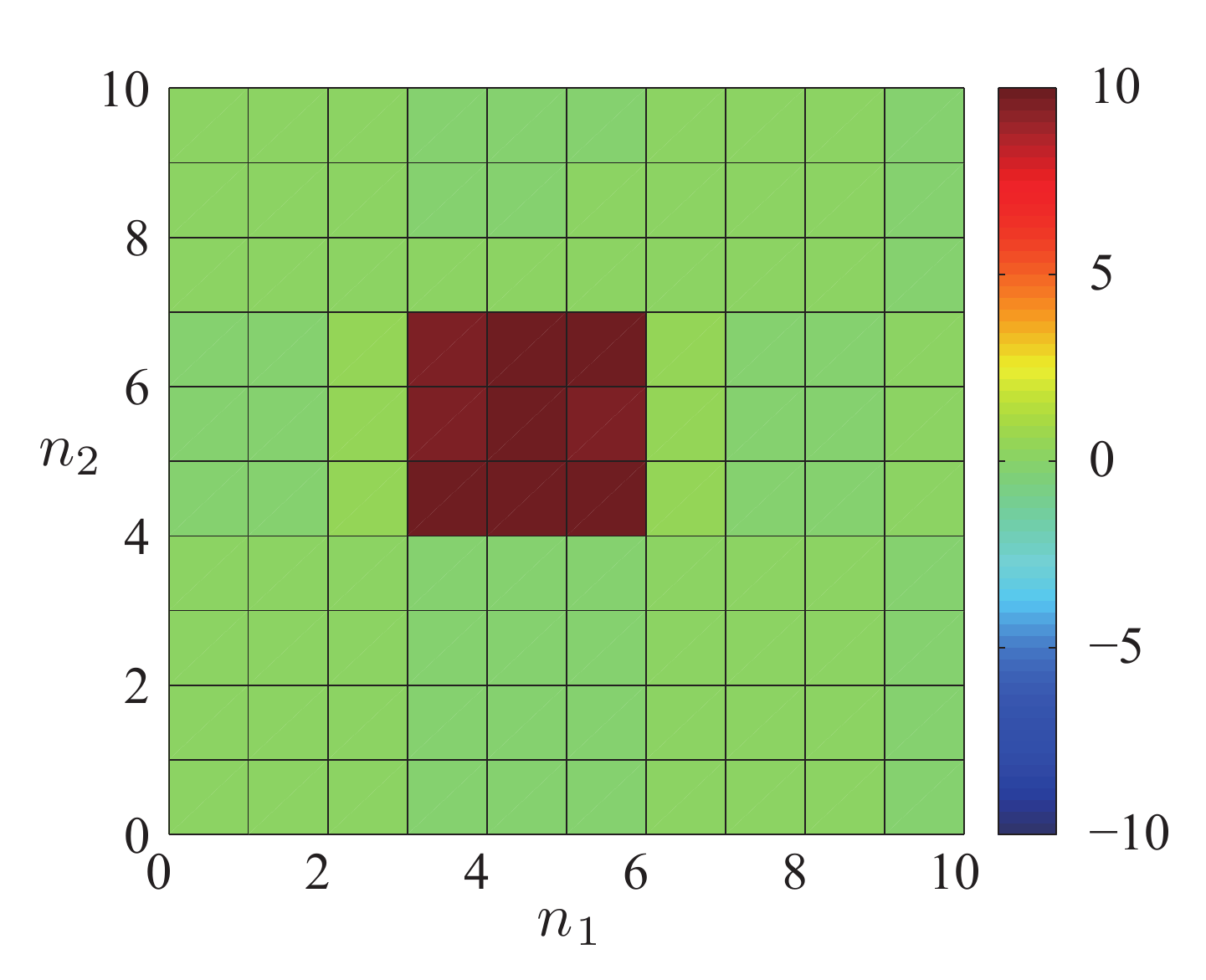} 
 \label{fig:2DAPG}
 }
 \caption{Examples of the estimated loss field
 tensor~($D = 2$, $M =
 60$, $\eta = 0$).}
 \label{fig:2Destimation}
\end{figure}
\begin{figure}[ttt]
 \centering
 \subfloat[Vector recovery]{\includegraphics[scale=0.61,clip]{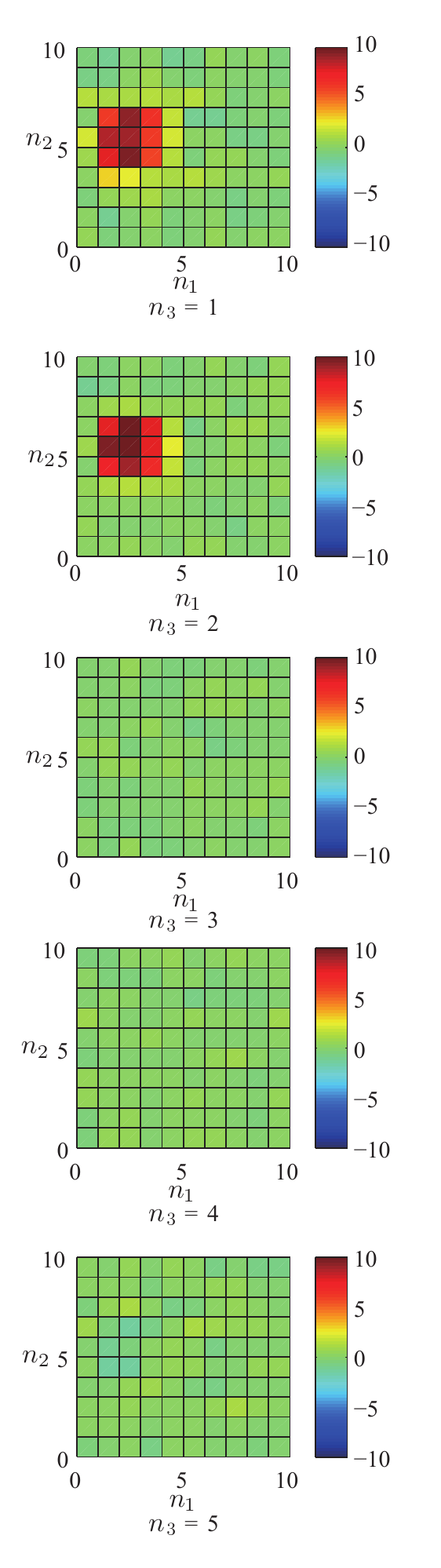}  \label{fig:3DDCT}}
  \subfloat[Tensor
 recovery]{\includegraphics[scale=0.64,clip]{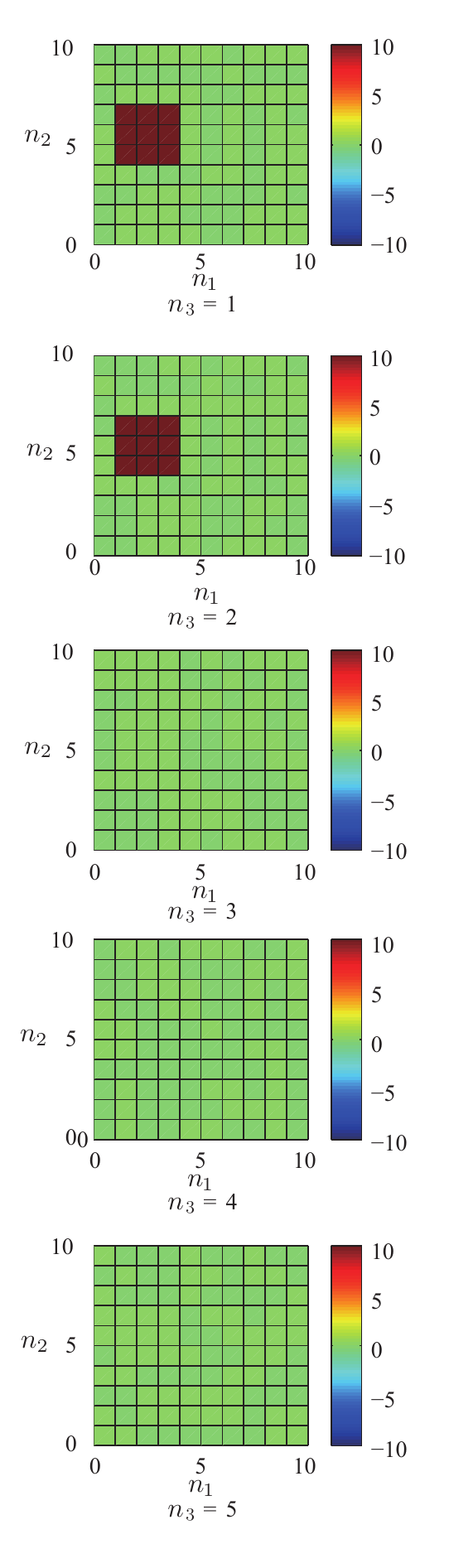}
 \label{fig:3DDR} 
 }
\caption{Examples of the estimated loss field tensor
using the vector and
 tensor recovery-based wireless tomography~~($D = 3$, $M =
 300$, $\eta = 0$).}
 \label{fig:3Destimation}
\end{figure}

\begin{figure*}[t]
\centering
 \subfloat[Vector
 recovery.]{\includegraphics[scale=0.55,clip]{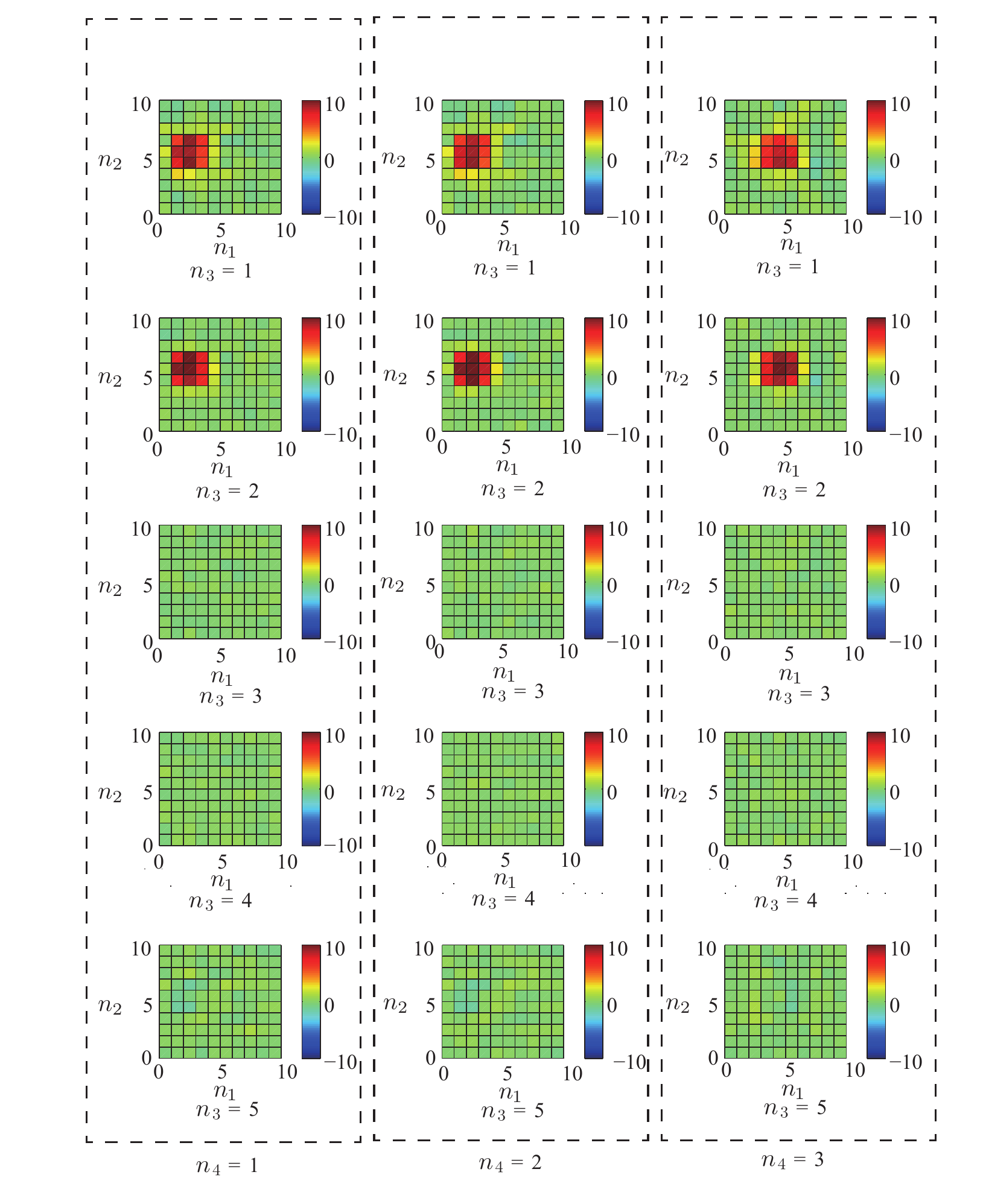} 
 \label{fig:4D_DCT_M300}}~~
\subfloat[Tensor
 recovery. ]{\includegraphics[scale=0.55,clip]{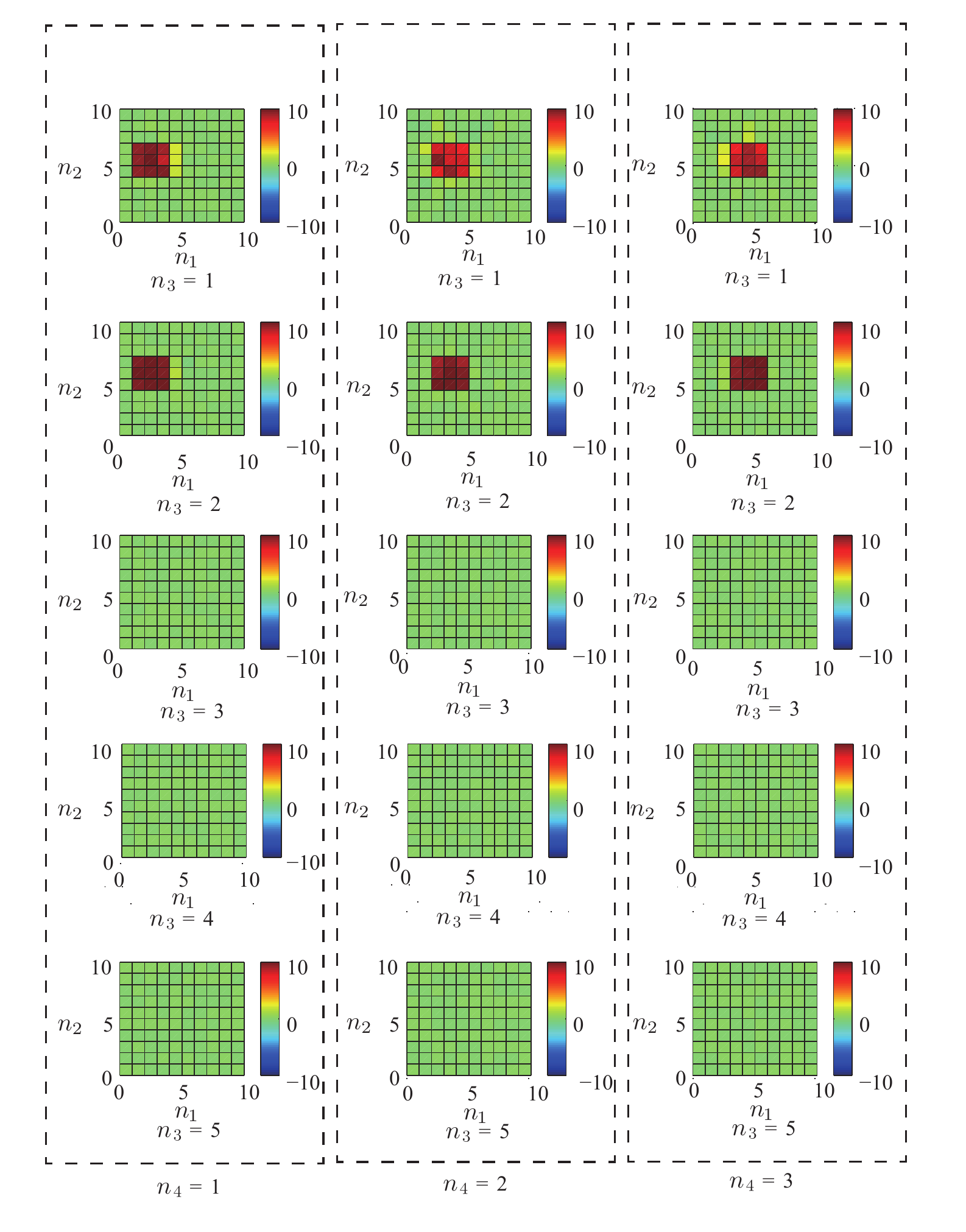} 
 \label{fig:4D_Tensor_M300}
}
 \caption{Examples of the estimated loss
 field tensor~($D = 4$, $M =
 900$, $M_{n_4} = 300$~$(n_4 = 1, 2, 3)$, $\eta = 0$).}
 \label{fig:4Destimation}
\end{figure*}

Figs.~\ref{fig:2Destimation}, \ref{fig:3Destimation}, and
\ref{fig:4Destimation} show examples of the estimated loss
field tensors
for $D = 2$, $3$, and $4$, respectively.
For each figure, we show the
loss field tensors estimated by the vector
and tensor recovery schemes in
the noiseless environment~(i.e., $\eta = 0$).
Here,
the number ($M$)
of measurements is set to $60$
for the case of $D= 2$, 
and $M$ is set to $300$ for the case of $D = 3$.
For the case of $D = 4$, furthermore,
$M$ is set to $900$,
and the number
$M_{n_4}$~$(n_4 = 1, 2, 3)$ of measurements
at time $n_4$ is
set to $300$.
From these figures, we observe that
the tensor
recovery scheme~(i.e., Figs.~\ref{fig:2DAPG}, \ref{fig:3DDR}, and
\ref{fig:4D_Tensor_M300}) can estimate the loss field tensor
more
accurately than
the vector recovery scheme~(i.e., Figs.~\ref{fig:2DDCT},
\ref{fig:3DDCT}, and \ref{fig:4D_DCT_M300}).

\begin{figure}[!t]
 \centering
 \includegraphics[scale=0.35]{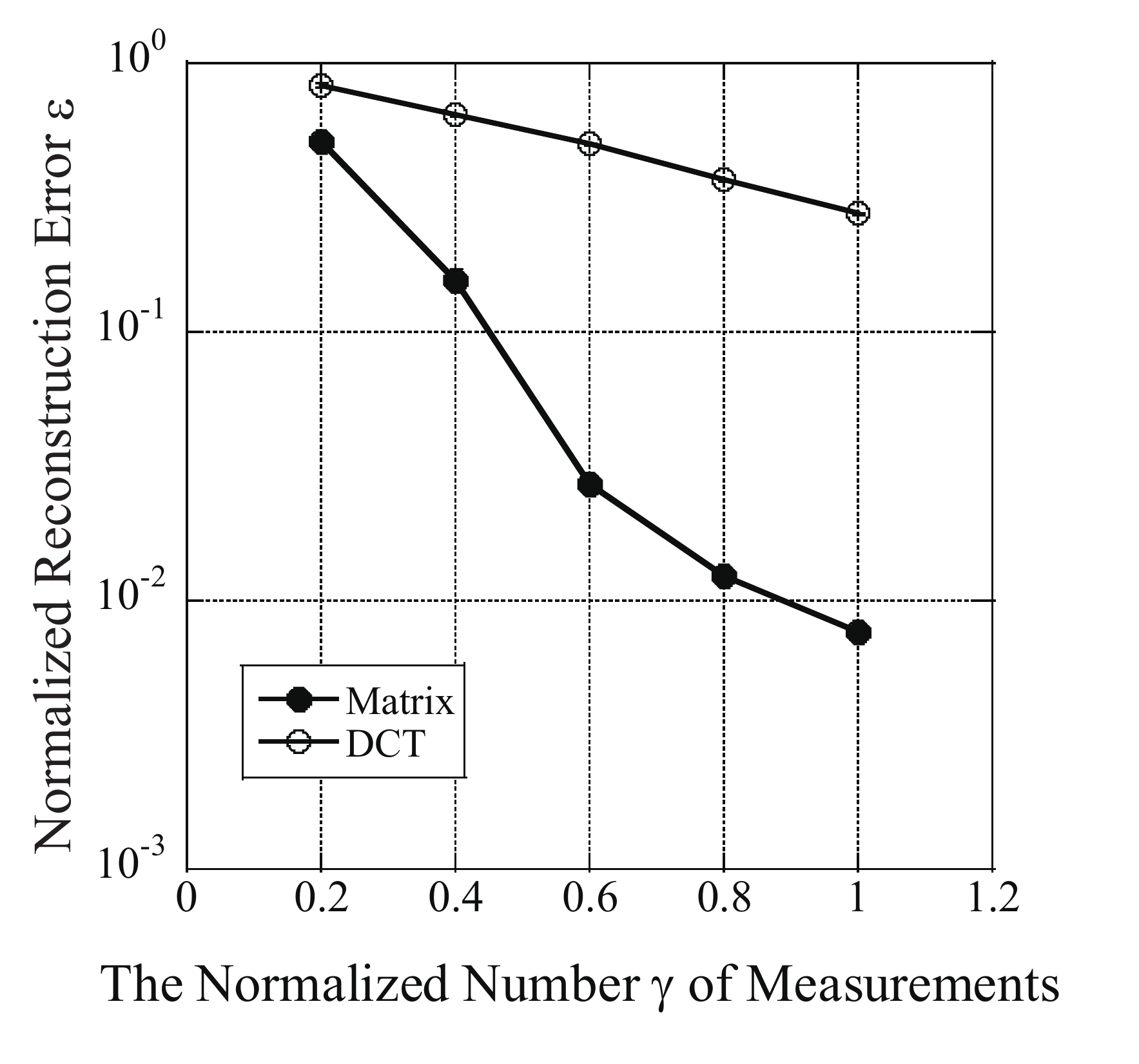}
 \caption{The normalized reconstruction error $\epsilon$ vs.\ the number
 $M$ of measurements~($D = 2$, $\eta = 0$).}
 \label{fig:error_vs_M_2D}
\end{figure}

\begin{figure}[!t]
 \centering
 \includegraphics[scale=0.35]{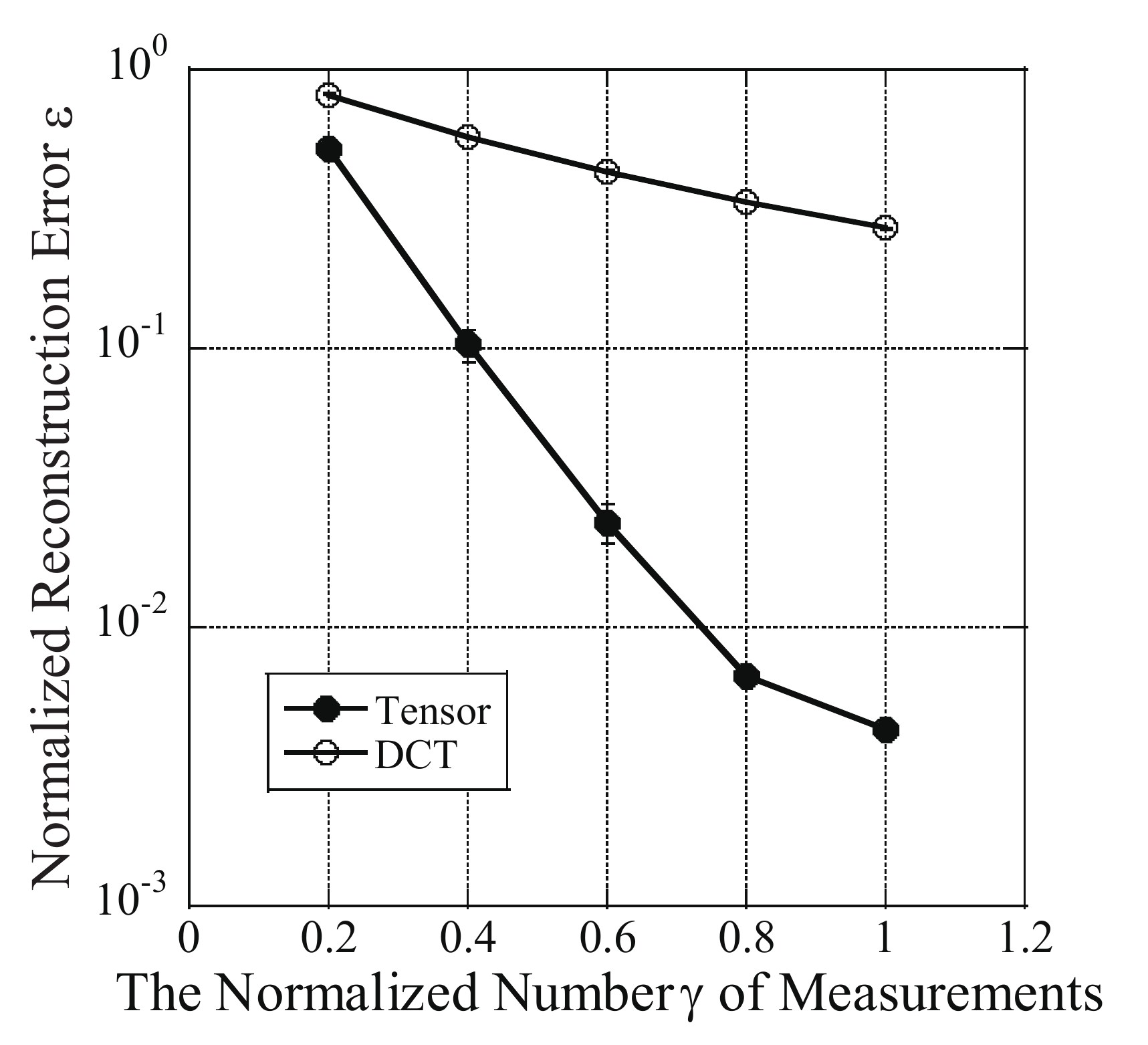}
 \caption{The normalized reconstruction error vs.\ the number $M$ of
 measurements~($D = 3$, $\eta = 0$).}
 \label{fig:error_vs_M_3D}
\end{figure}

\begin{figure}[!t]
 \centering
 \includegraphics[scale=0.35]{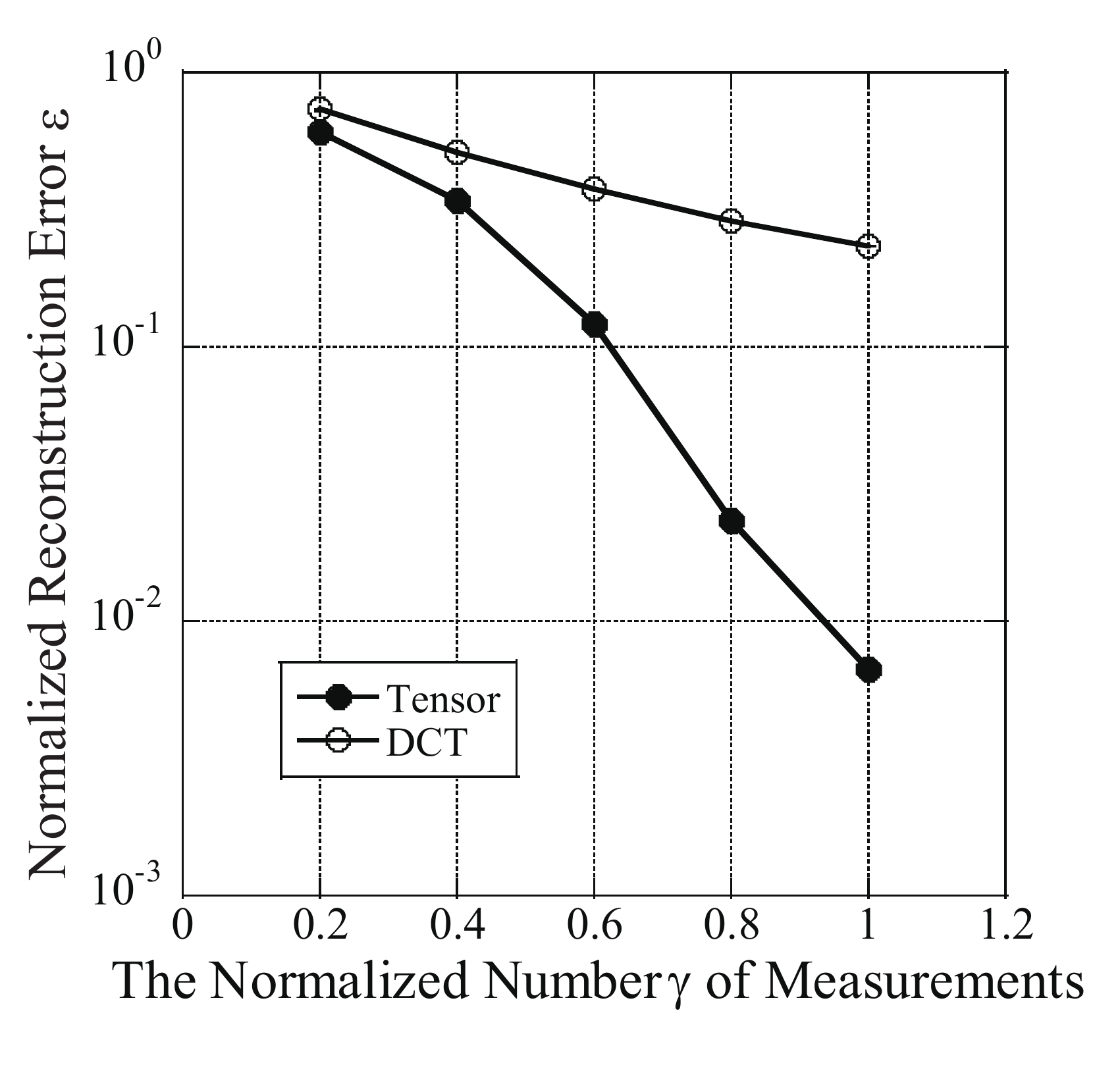}
 \caption{The normalized reconstruction error vs.\ the number $M$ of
 measurements~($M = 4$, $\eta = 0$).}
 \label{fig:error_vs_M_4D}
\end{figure}

For fair comparison
of the results obtained in
different dimensions,
we define the normalized number $\gamma$
of measurements as 
\[
 \gamma = \frac{M}{N_1N_2N_3N_4},
\] 
where $N_3 = N_4 = 1$ for $D = 2$ and $N_4 = 1$ for $D = 3$. 
Figs.~\ref{fig:error_vs_M_2D}, \ref{fig:error_vs_M_3D}, and
\ref{fig:error_vs_M_4D} show
the reconstruction error $\epsilon$ vs.\
$\gamma$ in the noiseless
environment~(i.e., $\eta = 0$) for $D = 2$, $3$, and $4$,
respectively.
In these figures, ``DCT'', ``Matrix'', and ``Tensor'' 
represent the performances of
the vector, matrix and tensor recovery schemes, respectively,
where we
set the number $N_{\rm run}$ of simulation experiments
for each parameter to $50$. 
From these figures, we observe that
the tensor recovery scheme
can achieve lower reconstruction errors than the vector
recovery scheme. 

\begin{figure}[!t]
 \centering
 \includegraphics[scale=0.35]{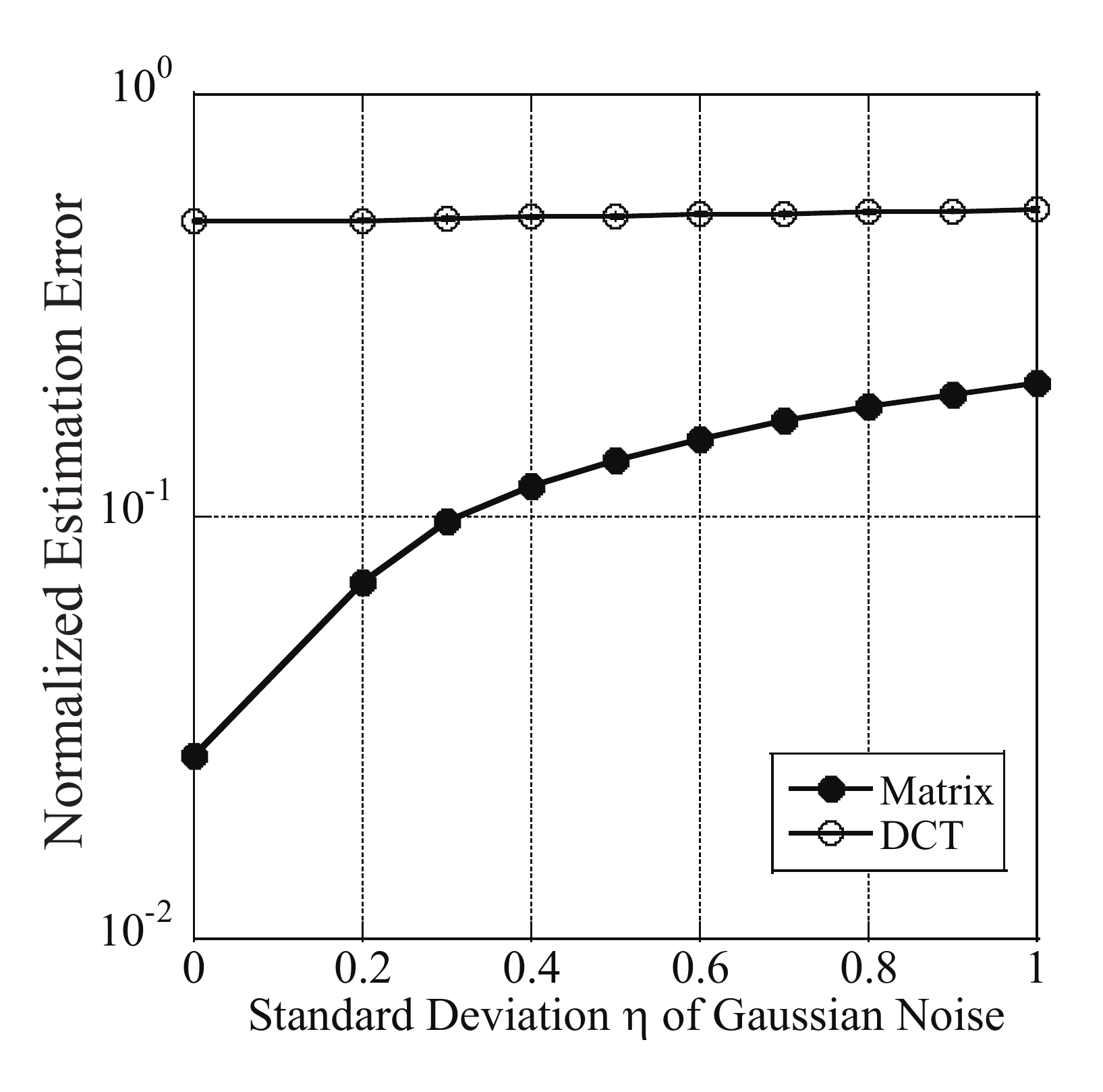}
 \caption{The normalized reconstruction error $\epsilon$ vs.\ standard
 deviation $\eta$ of noise~($D = 2$, $M = 60$).}
 \label{fig:error_vs_sigma_2D}
\end{figure}

\begin{figure}[!t]
 \centering
 \includegraphics[scale=0.35]{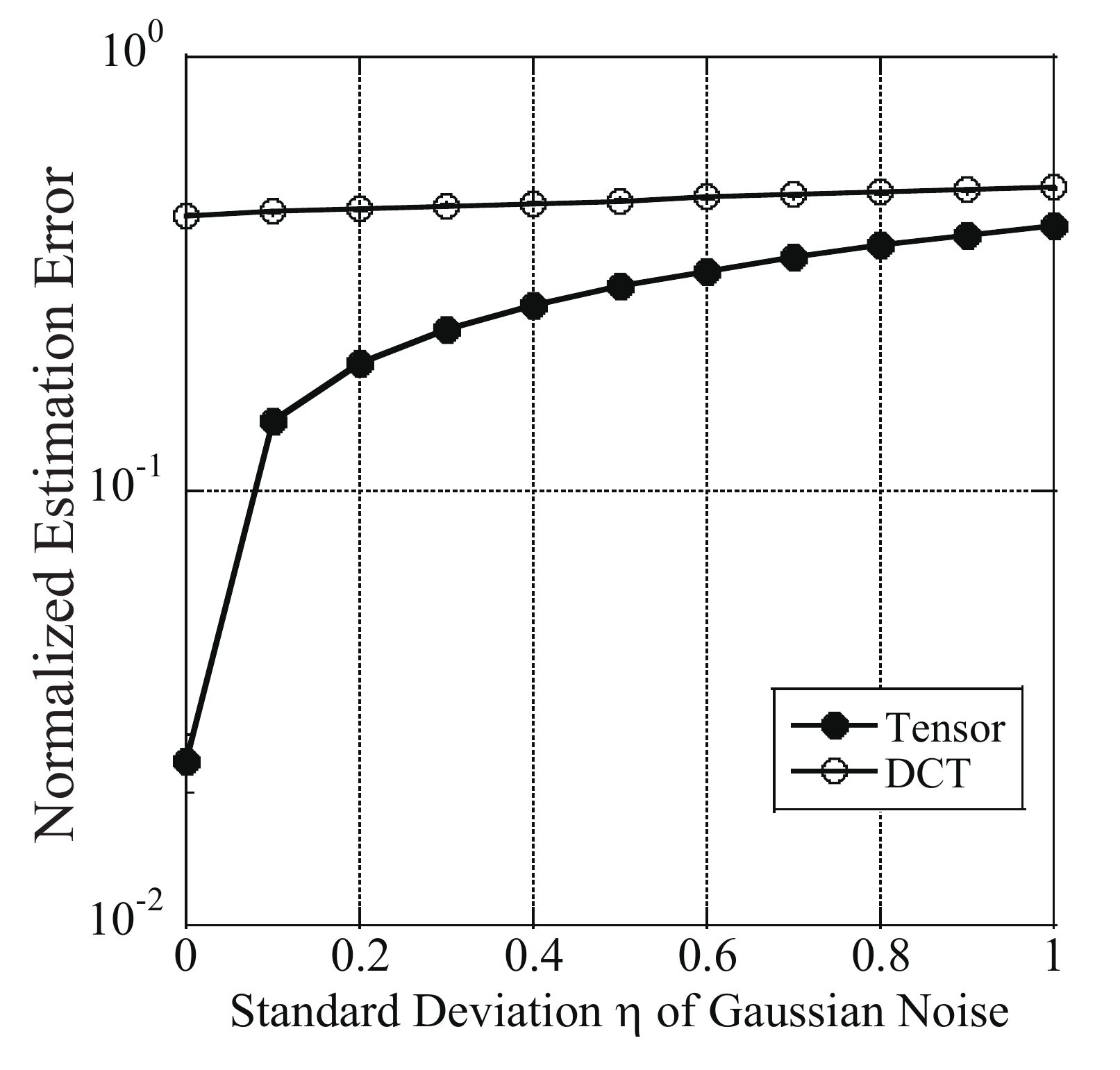}
 \caption{The normalized reconstruction error vs.\ standard deviation
 $\eta$ of noise~($D = 3$, $M = 300$).}
 \label{fig:error_vs_sigma_3D}
\end{figure}

\begin{figure}[!t]
 \centering
 \includegraphics[scale=0.35]{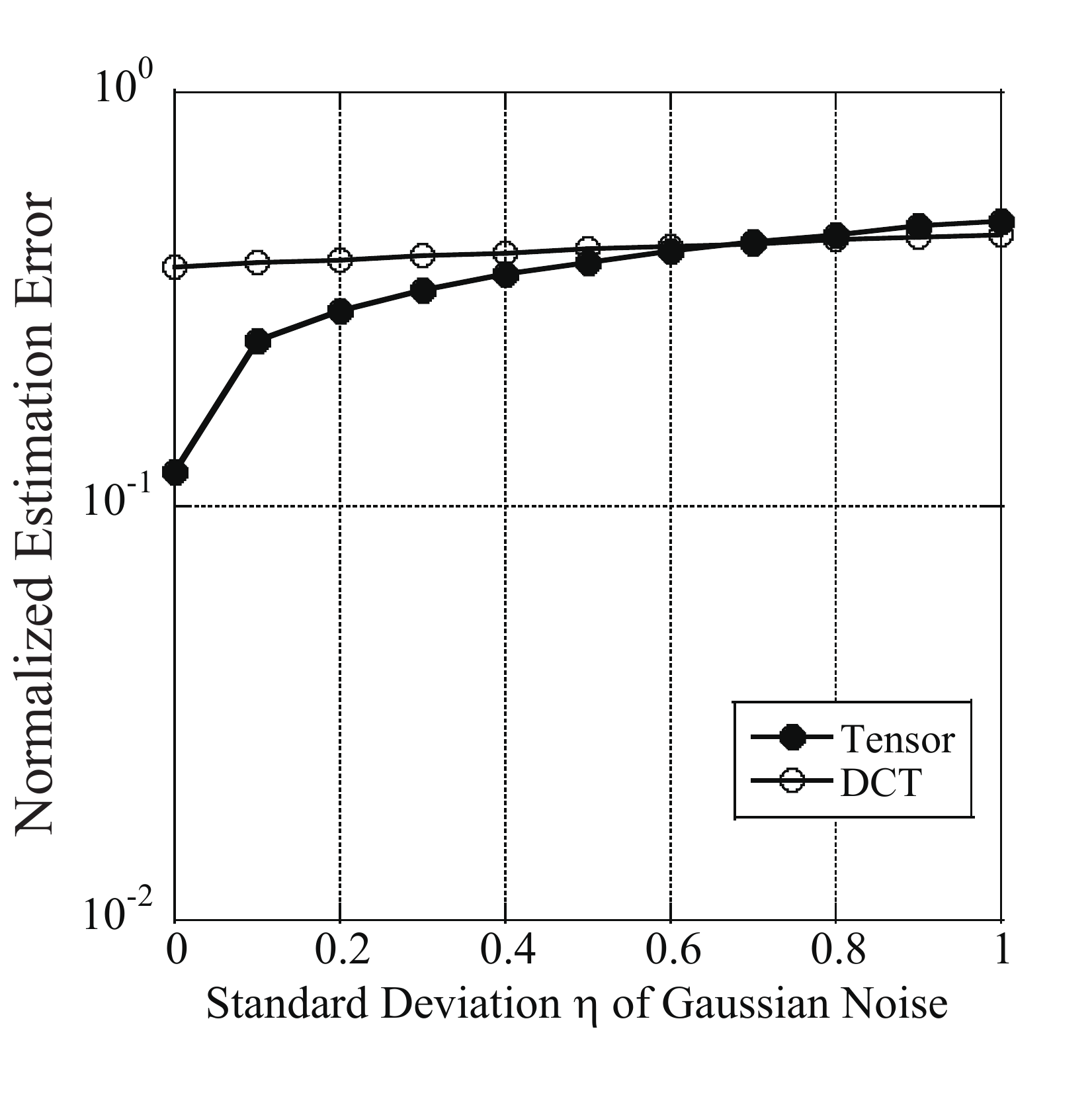}
 \caption{The normalized reconstruction error vs.\  standard deviation
 $\eta$ of noise~($D = 4$, $M = 900$).}
 \label{fig:error_vs_sigma_4D}
\end{figure}

Figs.~\ref{fig:error_vs_sigma_2D}, \ref{fig:error_vs_sigma_3D}, and
\ref{fig:error_vs_sigma_4D} show
the normalized reconstruction error
$\epsilon$ vs.\ the standard deviation $\eta$ of
noise for $D = 2$, $3$, and
$4$, respectively,
where we
set $M = 60$ for $D = 2$, $M = 300$ for $D = 3$,
and $M = 900$ and
$M_{n_4} = 300$~($n_4= 1, 2, 3$) for $D = 4$. 
For the cases of $D = 2$ and $3$, we observe that
the tensor recovery
scheme has lower reconstruction errors even
in the noisy environments. 
For the case of $D = 4$, 
however,
we observe that the tensor recovery scheme
achieves a better performance than
the vector recovery scheme only for smaller~$\eta$.
These figures indicate that the tensor
recovery scheme is
vulnerable to the measurement noise. 
Therefore, in order to ensure the performance
gain by the tensor recovery scheme,
measurement techniques with smaller
observation noise are required,
where we leave
it as a future work.

\subsection{Discussion: Vector Recovery vs.\ Tensor Recovery}
 \label{comparison}  
In this subsection, we discuss the reason
why the tensor recovery scheme can achieve
more accurate estimation of loss field tensors than
the vector recovery scheme, as
shown in the previous subsection. 
In order to simplify the discussion,
we focus on the case of $D = 2$.

In the vector recovery scheme, by using the frequency domain
representation $\bm{S}$, 
the loss field tensor $\bm{X}$ can be written as~\cite{Jain1989}
\[
 \bm{X} = 
 \bm{F}_{N_1}^{\rm \top} \bm{S} \bm{F}_{N_2}.
\]
When $\bm{X}$ has a high spatial
correlation,
most of the energy in $\bm{X}$ is concentrated in a few
low-frequency elements of $\bm{S}$, that is, $\bm{S}$ is an
approximately sparse matrix.
Therefore, when applying the vector recovery to
the measurement vector $\bm{y}$, we obtain
a sparsified matrix $\hat{\bm{S}}$ by
replacing the elements with a
smaller absolute value in $\bm{S}$ with zeros.
We thus obtain the estimated loss
field tensor $\hat{\bm{X}}^{\rm (V)}$ as
\begin{equation*}
  \hat{\bm{X}}^{\rm (V)} = \bm{F}_{N_1}^{\rm \top} \hat{\bm{S}}
   \bm{F}_{N_2}. 
\end{equation*}
The reconstruction error $\kappa_{\rm (V)}$ of $\hat{\bm{X}}^{\rm
(V)}$ is then given by
\[
  \kappa_{\rm (V)} = \|\hat{\bm{X}}^{\rm (V)}-\bm{X} \|_{\rm F}.
\]
\begin{figure}[t]
\centering
 \subfloat[Frequency representation of a loss field
 tensor]{\includegraphics[scale=0.35]{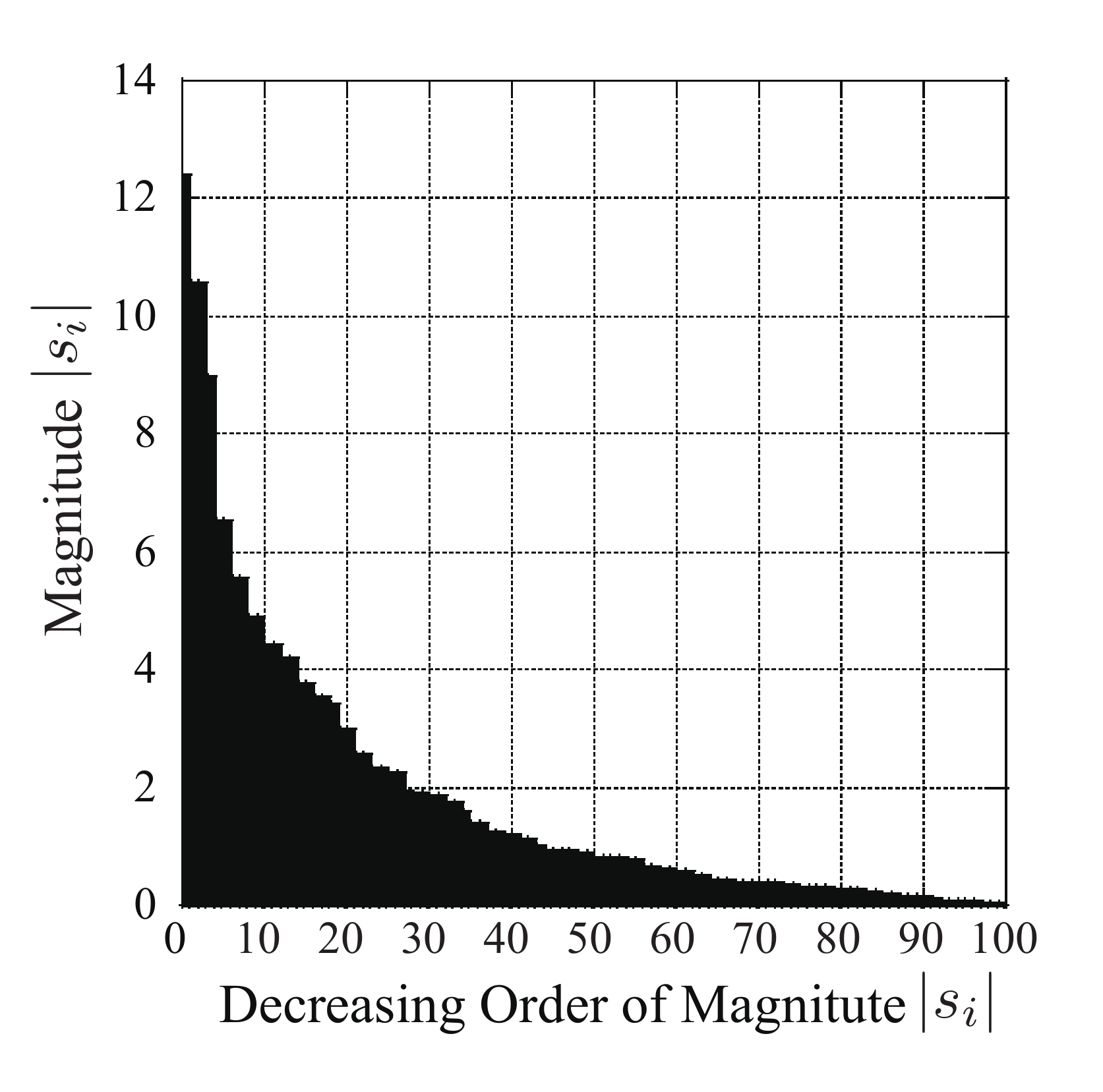}   
 \label{fig:true_spectrum}}\\
\subfloat[Frequency representation of a loss field
 Tensor estimated by the vector recovery scheme.]{\includegraphics[scale=0.35]{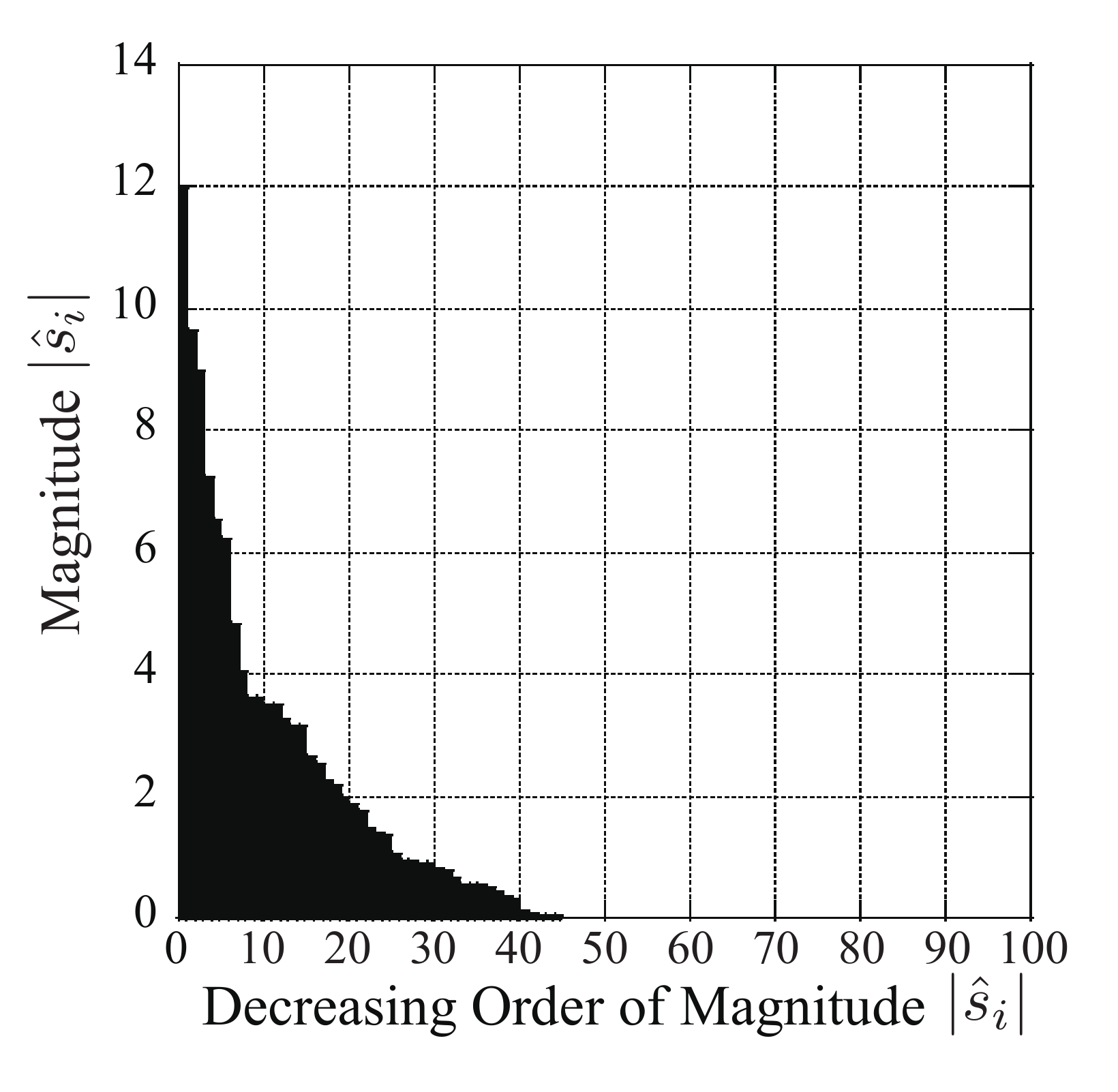} 
 \label{fig:estimated_spectrum}
}
 \caption{An example of ordered spectrum of a loss field tensor. }
 \label{fig:spectrum}
\end{figure}
Figs.~\ref{fig:true_spectrum} and
\ref{fig:estimated_spectrum} show $|s_i|$ and $|\hat{s}_i|$~$(i =
1, 2, \ldots, N_1N_2)$ sorted in the decreasing order,
where
$\bm{s} = \mathrm{vec}(\bm{S}) = (s_1~s_2~\cdots~s_{N_1N_2})^\top$
and $\hat{\bm{s}} = \mathrm{vec}({\hat{\bm{S}}}) =
(\hat{s}_1~\hat{s}_2~\cdots~\hat{s}_{N_1N_2})^\top$
are defined as the
vectorization of $\bm{S}$ and
$\hat{\bm{S}}$,
respectively.
We observe that
$\hat{\bm{s}}$ is obtained by
replacing the smaller elements of $\bm{s}$
with zeros.

On the other hand,
in the tensor recovery scheme,
the reconstruction
error can be explained with SVD. 
Namely, let us
define $\sigma_i\ (i=1,2,\ldots,r)$ as singular values of $\bm{X}$, where 
$r=\min\{N_1, N_2\}$.
With matrices $\bm{U} \in {\cal R}^{N_1 \times r}$ and 
$\bm{V} \in {\cal R}^{N_2 \times r}$ whose column vectors are orthogonal, 
the SVD of $\bm{X}$ can be written as
\begin{equation*}
 \bm{X}=\bm{U \Sigma V}^{\top},
\end{equation*}
where 
$\bm{\Sigma}={\rm diag}(\sigma_1,\sigma_2,\ldots,\sigma_r)$ denotes an
$r \times r$ diagonal matrix.

Without loss of generality, we can assume that singular values 
$\sigma_i\ (i=1,2,\ldots,r)$ are arranged in the decreasing order, i.e.,  
$\sigma_1 \geq \sigma_2 \geq \cdots \geq \sigma_r \geq 0$.
Applying the tensor recovery scheme to the
measurement vector $\bm{y}$, we obtain a diagonal matrix
$\hat{\bm{\Sigma}} = \mathrm{diag}(\hat{\sigma}_1, \hat{\sigma}_2,
\ldots, \hat{\sigma}_r)$ by replacing the
$r - K$ smaller diagonal elements
of $\bm{\Sigma}$ with zeros, that is, $\hat{\sigma}_i \approx
\sigma_i$~$(i = 1, 2, \ldots, K)$,
$\hat{\sigma}_{K+1}=\hat{\sigma}_{K+2}= \cdots 
= \hat{\sigma}_r =0$. We thus obtain estimated loss field tensor
$\hat{\bm{X}}^{(\rm M)}$ as
\begin{equation*}
 \hat{\bm{X}}^{\rm (M)}=\bm{U}\hat{\bm{\Sigma}}\bm{V}^{\top}.
\end{equation*}
The reconstruction error $\kappa_{\rm (M)}$  of $\hat{\bm{X}}^{(\rm M)}$
is then given by
\begin{equation*}
 \kappa_{\rm (M)} = \|\hat{\bm{X}}^{\rm (M)}-\bm{X} \|_{\rm F}.
\end{equation*}

Table~\ref{tab:svd} shows singular values of the true loss field tensor and an
estimated loss field tensor.
Because the rank of the
loss field tensor $\bm{X}$ in Fig.~\ref{fig:channel2D} is
$\mathrm{rank}\bm{X} = 1$,
$\bm{X}$ has only one non-zero singular value
$\sigma_1$.
The estimated loss field tensor $\hat{\bm{X}}$ highly
approximates $\bm{X}$
by replacing $\sigma_i$~($i = 3, 4, \ldots,
N_1N_2$) with zeros. 
\begin{table}[t]
 \centering
 \caption{Singular values of a loss field tensor and its estimated loss
 field tensor~($D = 2$, $M = 60$).}
\label{tab:svd}
 \begin{tabular}[t]{|c|c|c|c|c|c|}
  \hline
  $i$ & $1$ & $2$ & $3, 4, \ldots, N_1N_2$ \\
  \hline
  $\sigma_i$ & $30.00$ & $0$ & $0$ \\
  \hline
  $\hat{\sigma}_i$ & $29.68$& $0.001244$ &  $0$\\
  \hline
 \end{tabular}
\end{table}

Finally,
suppose that both $\hat{\bm{S}}$ and $\hat{\bm{\Sigma}}$
generally have
$K$ non-zero elements.
For the case of $D = 2$, it is well-known that
SVD provides
the smallest reconstruction error, that is, $\kappa_{\rm (M)} \leq
\kappa_{\rm (V)}$~\cite{Jain1989,Dapena2002}, where  
$\kappa_{\rm (M)}$ is given by 
\begin{equation*}
 \kappa_{\rm (M)} = \sqrt{\sum_{i=K+1}^{r}\sigma_i^2}. 
\end{equation*}
Therefore, 
for the case of $D = 2$, 
the tensor recovery scheme is the best way in terms of the
reconstruction error. 
Actually, in~\cite{Dapena2002}, the authors show
that the SVD-based image
compression obtains better reconstruction errors than
the DCT-based
image compression. 
Furthermore, in~\cite{Lathauwer2000},
HOSVD is studied and it is shown that
low-rank approximation of the $D$-th tensor
provides a good approximation
in terms of the reconstruction error.

\section{Conclusion}
\label{sec:conclusion}
In this paper, we proposed a multi-dimensional
wireless tomography using
tensor-based compressed sensing,
which enables us to estimate the
locations of internal obstructions
by a small number of
measurement signals.
While the conventional wireless tomography using
the vector recovery-based compressed sensing utilizes
the sparsity of the
frequency representation of a given loss field tensor,
the proposed
wireless tomography utilizes its low-rank property.
With simulation
experiments, we validated the effects of the proposed wireless
tomography, and showed that the tensor
recovery-based wireless tomography
can provide more
accurate estimation of the loss field tensor,
especially in a less noisy
environment. 

We still have some remaining issues to be resolved.
For example, in
order to achieve noiseless measurements,
we require a signal processing
technique to extract a direct link in multipath fading
environments.
In this paper, we assumed that many wireless nodes
are deployed on the border of the monitored region,
and we chose
wireless links randomly by using these wireless nodes.
In a situation
where there are a small number of wireless nodes, however,
we have to consider a
wireless node selection scheme to achieve an efficient
estimation of a loss
field tensor. We will try these issues in the future work.


%





\ifCLASSOPTIONcaptionsoff
  \newpage
\fi

\end{document}